\documentclass[10pt,conference]{IEEEtran}
\IEEEoverridecommandlockouts

\usepackage[switch]{lineno}
\usepackage{amsmath}
\usepackage{amssymb}
\usepackage{amsfonts}
\usepackage{booktabs}
\usepackage{comment}
\usepackage{url}
\usepackage{listings}
\usepackage{caption}
\usepackage{subcaption}
\usepackage[htt]{hyphenat}
\usepackage{xcolor,colortbl}
\usepackage[ruled, linesnumbered]{algorithm2e}
\usepackage{tabularx}
\usepackage{makecell}
\usepackage[]{soul}
\usepackage[]{todonotes}
\usepackage{xspace}
\usepackage[nomessages]{fp}
\usepackage[autolanguage]{numprint}		%
\usepackage[framemethod=tikz]{mdframed}	%
\usepackage[subtle]{tex/savetrees}
\usepackage{microtype}

\mdfdefinestyle{graybox}{
	roundcorner=1mm,%
	backgroundcolor=gray!30
}

\makeatletter
\let\old@lstKV@SwitchCases\lstKV@SwitchCases
\def\lstKV@SwitchCases#1#2#3{}
\makeatother
\usepackage{lstlinebgrd}
\makeatletter
\let\lstKV@SwitchCases\old@lstKV@SwitchCases

\lst@Key{numbers}{none}{%
	\def\lst@PlaceNumber{\lst@linebgrd}%
	\lstKV@SwitchCases{#1}%
	{none:\\%
		left:\def\lst@PlaceNumber{\llap{\normalfont
				\lst@numberstyle{\thelstnumber}\kern\lst@numbersep}\lst@linebgrd}\\%
		right:\def\lst@PlaceNumber{\rlap{\normalfont
				\kern\linewidth \kern\lst@numbersep
				\lst@numberstyle{\thelstnumber}}\lst@linebgrd}%
	}{\PackageError{Listings}{Numbers #1 unknown}\@ehc}}
\makeatother

\graphicspath{{figures/}}

\tikzset{inlinenotestyle/.append style={align=justify}}

\SetNoFillComment
\DontPrintSemicolon

\SetCommentSty{mycommfont}
\SetFuncSty{texttt}

\newcolumntype{C}[1]{>{\centering\arraybackslash}p{#1}}
\newcolumntype{Y}{>{\raggedleft\arraybackslash}X}
\newcolumntype{Z}{>{\centering\arraybackslash}X}

\newenvironment{conditions*}
{\par\vspace{\abovedisplayskip}\noindent
	\tabularx{\columnwidth}{>{$}l<{$} @{${}={}$} >{\raggedright\arraybackslash}X}}
{\endtabularx\par\vspace{\belowdisplayskip}}

\lstdefinelanguage{llvm}{
	morecomment = [l]{;},
	morecomment=[l]{//},
	morestring=[b]", 
	sensitive = true,
	classoffset=0,
	commentstyle=\itshape\color{blue!100!black},       %
	morekeywords={
		define, declare, global, constant,
		internal, external, private,
		linkonce, linkonce_odr, weak, weak_odr, appending,
		common, extern_weak,
		thread_local, dllimport, dllexport,
		hidden, protected, default,
		except, deplibs,
		volatile, fastcc, coldcc, cc, ccc,
		x86_stdcallcc, x86_fastcallcc,
		ptx_kernel, ptx_device,
		signext, zeroext, inreg, sret, nounwind, noreturn,
		nocapture, byval, nest, readnone, readonly, noalias, uwtable,
		inlinehint, noinline, alwaysinline, optsize, ssp, sspreq,
		noredzone, noimplicitfloat, naked, alignstack,
		module, asm, align, tail, to,
		addrspace, section, alias, sideeffect, c, gc,
		target, datalayout, triple,
		blockaddress,
		@malloc,@strncpy,@tmpfile,@foo
	},
	classoffset=1, keywordstyle=\color{purple},
	morekeywords={
		fadd, sub, fsub, mul, fmul,
		sdiv, udiv, fdiv, srem, urem, frem,
		and, or, xor,
		icmp, fcmp,
		eq, ne, ugt, uge, ult, ule, sgt, sge, slt, sle,
		oeq, ogt, oge, olt, ole, one, ord, ueq, ugt, uge,
		ult, ule, une, uno,
		nuw, nsw, exact, inbounds,
		phi, call, select, shl, lshr, ashr, va_arg,
		trunc, zext, sext,
		fptrunc, fpext, fptoui, fptosi, uitofp, sitofp,
		ptrtoint, inttoptr, bitcast,
		ret, br, indirectbr, switch, invoke, unwind, unreachable,
		malloc, alloca, free, load, store, getelementptr,
		extractelement, insertelement, shufflevector,
		extractvalue, insertvalue,
	},
	alsoletter={\%},
}

\newcommand{\defnote}[3]{
	\newcounter{#1}
	
	\expandafter
	\newcommand\csname cmt#1\endcsname[1]{
		\refstepcounter{#1} 
		\textcolor{#3}{\textbf{#2 [\csname the#1\endcsname]:}} %
		\hl{\textbf{##1.}}} %
	
	\expandafter
	\newcommand\csname cmtdel#1\endcsname[2]{
		\refstepcounter{#1} 
		\textcolor{#3}{\textbf{#2 [\csname the#1\endcsname]:} ##2} %
		\st{\textbf{##1}}} %
	
	\expandafter
	\newcommand\csname #1\endcsname[1]{%
		\refstepcounter{#1}%
		{%
			\todo[color={#3!30},inline]{%
				\textbf{#2 [\csname the#1\endcsname]:}~##1}%
	}}
}

\defnote{dd}{Dipanjan}{magenta}
\defnote{pb}{Priyanka}{blue}
\defnote{ac}{Andrea}{caribbeangreen}
\defnote{ab}{Antonio}{cyan}
\defnote{ig}{Ilya}{emerald}
\defnote{sv}{Saastha}{arylideyellow}

\newcommand{\etal}{\textit{et. al.}}
\newcommand{\etc}{\textit{etc.}}
\newcommand{\eg}{\textit{e.g.}}
\newcommand{\viz}{\textit{viz.}}
\newcommand{\ie}{\textit{i.e.}}

\newcommand{\Ln}[1]{\ensuremath{\sf Line~ #1}}
\newcommand{\Lns}[2]{\ensuremath{\sf Lines~ #1--#2}}
\newcommand{\Tbl}[1]{\ensuremath{\sf Table~\ref{#1}}}
\newcommand{\Fig}[1]{\ensuremath{\sf Figure~\ref{#1}}}
\newcommand{\Sec}[1]{\ensuremath{\sf Section~\ref{#1}}}

\newcommand{\Alg}[1]{\ensuremath{\sf Algorithm~\ref{#1}}}

\let\num\numprint

\definecolor{arylideyellow}{rgb}{0.91, 0.84, 0.42}
\definecolor{caribbeangreen}{rgb}{0.0, 0.8, 0.6}
\definecolor{emerald}{rgb}{0.31, 0.78, 0.47}

\newcommand\crule[3][black]{\textcolor{#1}{\rule{#2}{#3}}}

\SetKwFunction{getsubclasses}{GetSubClasses}
\SetKwFunction{getsuperclasses}{GetSuperClasses}
\SetKwFunction{getclassmethods}{GetClassMethods}
\SetKwFunction{getmethodargtypes}{GetMethodArgTypes}
\SetKwFunction{getmethodsC}{GetMethodsFromClass}
\SetKwFunction{getmethodsa}{GetMethodsFromApk}
\SetKwFunction{getclassesj}{GetClassesFromJar}
\SetKwFunction{getclassesa}{GetClassesFromApk}
\SetKwFunction{getmethodname}{GetMethodName}
\SetKwFunction{iscompatible}{IsCompatible}
\SetKwFunction{getcg}{GetCallGraph}
\SetKwFunction{getch}{GetClassHierarchy}
\SetKwFunction{getclassdefiningmethod}{GetclassDefiningMethod}
\SetKwFunction{getmethodacc}{IsPublicOrProtected}
\SetKwFunction{getcallers}{GetCallers}
\SetKwFunction{fanalysis}{AndroidFrameworkAnalysis}
\SetKwFunction{aanalysis}{AppAnalysis}
\SetKw{kwextends}{ extends }
\SetKw{kwimplements}{ implements }
\SetKw{kwor}{ or }
\SetKwProg{Fn}{Function}{}{}
\SetKwInOut{Input}{Input}
\SetKwInOut{Output}{Output}

\SetKwFunction{callbackdep}{InterCallbackDependencies}
\SetKwFunction{classrefvariables}{GetClassDefinedRefVariables}
\SetKwFunction{updateweights}{UpdateWeights}
\SetKwFunction{updatepenalizeweights}{UpdateAndPenalizeWeights}
\SetKwFunction{decreaserankmin}{DecreaseRankMin}
\SetKwFunction{getcallbackargs}{GetCallbackArguments}
\SetKwFunction{isnullchecked}{IsNullChecked}
\SetKwFunction{fuzzcallback}{ExecuteCallback}
\SetKwFunction{fuzzexplore}{CallbackExploration}
\SetKwFunction{remove}{RemoveActivity}
\SetKwFunction{sizeof}{SizeOf}
\SetKwFunction{fuzz}{Fuzz}
\SetKwFunction{getforegroundact}{getLiveActivity}
\SetKwFunction{getnewclasses}{getNewClasses}
\SetKwFunction{getnextcallback}{getNextCallback}
\SetKwFunction{getnextclass}{getNextClass}
\SetKwFunction{generatearguments}{generateArguments}
\SetKwFunction{getinstance}{getInstance}
\SetKwFunction{spawnapp}{spawnApp}
\SetKwFunction{restartapp}{restartApp}
\SetKw{kwextends}{ extends }
\SetKw{kwimplements}{ implements }
\SetKw{kwreads}{ reads }
\SetKw{kwwrites}{ writes }
\SetKwProg{Fn}{Function}{}{}
\SetKwInOut{Input}{Input}
\SetKwInOut{Output}{Output}

\newcommand{\tool}{\textsc{Columbus}\xspace}
\newcommand{\android}{Android\xspace}
\newcommand{\activity}{activity\xspace}
\newcommand{\mypar}[1]{\vspace{0.25mm}\noindent\textbf{#1}\xspace}
\newcommand{\code}[1]{\texttt{#1}\xspace}
\newcommand{\monkey}{{\sc Monkey}\xspace}
\newcommand{\ape}{{\sc Ape}\xspace}
\newcommand{\stoat}{{\sc Stoat}\xspace}
\newcommand{\timemachine}{{\sc TimeMachine}\xspace}
\newcommand{\ehbdroid}{{\sc EHBDroid}\xspace}
\newcommand{\dynodroid}{{\sc Dynodroid}\xspace}
\newcommand{\acteve}{{\sc ACTEve}\xspace}

\newcommand{\puma}{{\sc PUMA}\xspace}
\newcommand{\npe}{\code{NullPointerException}}
\newcommand{\ise}{\code{IllegalStateException}}
\newcommand{\angr}{{\sc angr}\xspace}
\newcommand{\frida}{{\sc Frida}\xspace}

\newcommand{\emma}{{\sc Emma}\xspace}

\newcommand{\logcat}{{\sc Logcat}\xspace}
\newcommand{\androtest}{AndroTest\xspace}
\newcommand{\modulecallbackdiscovery}{\textit{callback discovery}\xspace}
\newcommand{\moduleargumentgeneration}{\textit{argument generation}\xspace}
\newcommand{\modulecallbackdependency}{\textit{callback dependency}\xspace}
\newcommand{\moduleexploration}{\textit{exploration}\xspace}
\newcommand{\feedbackdatadependency}{\textit{data dependency}\xspace}
\newcommand{\feedbackcrashguidance}{\textit{crash-guidance}\xspace}

\newcommand{\numandrotestapps}{68}
\newcommand{\numrealworldapps}{140}

\newcommand{\minrating}{4.5}
\newcommand{\mininstalls}{500000}
\newcommand{\testduration}{3}	%

\newcommand{\numrepetitions}{5}	%
\newcommand{\numrealworldappcategories}{14}

\newcommand{\numframeworkcallbacks}{30682}
\newcommand{\numappcallbacks}{4991}

\newcommand{\numcallbackswithatleastoneprimitiveargument}{1566}
\newcommand{\argumentgenerationsucceeded}{1332}
\newcommand{\argumentgenerationtimedout}{234}
\newcommand{\symbolicexecutiontimeout}{5}	%
\newcommand{\numreferencetypecallbacks}{4147}
\newcommand{\totalreferencetypeargs}{4857}
\newcommand{\numreferenceargsonheap}{4650}
\newcommand{\numreferenceargscreated}{207}

\newcommand{\totaldependencyrelations}{2456}
\newcommand{\totaldependencyonvariables}{975}

\newcommand{\numandrotestappsworked}{60}
\newcommand{\numandrotestappsdidnotwork}{8}
\newcommand{\numappslessthanonek}{30}
\newcommand{\numappsgreaterthanoneklessthanthreek}{17}
\newcommand{\numappsgreaterthanthreek}{13}

\newcommand{\numappswithcoveragemorethanstoat}{45}

\newcommand{\numappswithcoveragemorethanehbdroid}{55}

\newcommand{\numappswithcoveragemorethanape}{41}

\newcommand{\numappswithcoveragemorethantimemachine}{41}

\newcommand{\stoatcoveragebestinapps}{5}
\newcommand{\ehbdroidcoveragebestinapps}{2}
\newcommand{\apecoveragebestinapps}{10}
\newcommand{\timemachinecoveragebestinapps}{16}
\newcommand{\columbuscoveragebestinapps}{36}

\newcommand{\stoataveragecoverage}{46}
\newcommand{\ehbdroidaveragecoverage}{27}
\newcommand{\apeaveragecoverage}{53}
\newcommand{\timemachineaveragecoverage}{52}
\newcommand{\columbusaveragecoverage}{58}

\newcommand{\numappsanalyzedwherecolumbusisworse}{10}
\newcommand{\timeaftercolumbuscoverageexceedsothers}{5}		%
\newcommand{\timeuntilcoverageincreasesfast}{20}		%

\newcommand{\typesofandrotestcrash}{16}
\newcommand{\totalandrotestcrash}{137}
\newcommand{\numandrotestappswithcrash}{49}
\newcommand{\totalandrotestcrashwithFP}{153}

\newcommand{\numcrashesstoat}{31}
\newcommand{\numcrashesehbdroid}{25}
\newcommand{\numcrashesape}{111}
\newcommand{\numcrashestimemachine}{87}
\newcommand{\numcrashescolumbus}{137}
\newcommand{\numcrashescolumbuswithFP}{153}

\newcommand{\stoatcrashesbestinapps}{14}
\newcommand{\ehbdroidcrashesbestinapps}{10}
\newcommand{\apecrashesbestinapps}{25}
\newcommand{\timemachinecrashesbestinapps}{21}
\newcommand{\columbuscrashesbestinapps}{45}

\newcommand{\numcolumbusonlycrash}{55}
\newcommand{\numcolumbusonlycrashnotreproduced}{16}
\newcommand{\numappcallbackswithsetenabled}{71}
\newcommand{\numappcallbackswithsetenablednonlifecycle}{4}
\newcommand{\numappcallbacksinvokedwithreflection}{207}

\newcommand{\typesofrealworldcrash}{9}
\newcommand{\totalrealworldcrash}{70}
\newcommand{\numrealworldappswithcrash}{54}
\newcommand{\nullpointerexceptionsinrealworldapps}{22}
\newcommand{\illegalstateexceptionsinrealworldapps}{26}

\newcommand{\numcrashescolumbusdependency}{140}
\newcommand{\numappsaffectedbydependencyfeedback}{5}

\usepackage{fancyhdr}

\AtBeginDocument{%
  \providecommand\BibTeX{{%
    \normalfont B\kern-0.5em{\scshape i\kern-0.25em b}\kern-0.8em\TeX}}}

\begin{document}
\date{}
\title{\tool: Android App Testing Through Systematic Callback Exploration}

	\author{ \IEEEauthorblockN{ Priyanka Bose\IEEEauthorrefmark{1},
			Dipanjan Das\IEEEauthorrefmark{1},
			Saastha Vasan\IEEEauthorrefmark{1},
			Sebastiano Mariani\IEEEauthorrefmark{2},
			Ilya Grishchenko\IEEEauthorrefmark{1},
			Andrea Continella\IEEEauthorrefmark{3},\\
			Antonio Bianchi\IEEEauthorrefmark{4},
			Christopher Kruegel\IEEEauthorrefmark{1}, and
			Giovanni Vigna\IEEEauthorrefmark{1}}
		
			\IEEEauthorblockA{\IEEEauthorrefmark{1}University of California, Santa Barbara \{priyanka,dipanjan,saastha,grishchenko,chris,vigna\}@cs.ucsb.edu}
			\IEEEauthorblockA{\IEEEauthorrefmark{2}VMware, Inc.
			\{smariani\}@vmware.com}\IEEEauthorblockA{\IEEEauthorrefmark{3}University of Twente
			\{a.continella\}@utwente.nl}
			\IEEEauthorblockA{\IEEEauthorrefmark{4}Purdue University
				\{antoniob\}@purdue.edu}
			}

\microtypesetup{disable}
\maketitle

\thispagestyle{fancy}

\begin{abstract}
With the continuous rise in the popularity of \android mobile devices, automated testing of apps has become more important than ever.
\android apps are event-driven programs.
Unfortunately, generating all possible types of events by interacting with an app's interface is challenging for an automated testing approach.
Callback-driven testing eliminates the need for event generation by directly invoking app callbacks.
However, existing callback-driven testing techniques assume prior knowledge of \android callbacks, and they rely on a human expert, who is familiar with the \android API, to write \textit{stub} code that prepares callback arguments before invocation.
Since the \android API is very large and keeps evolving, prior techniques could only support a small fraction of callbacks present in the \android framework.

In this work, we introduce \tool, a callback-driven testing technique that employs two strategies to eliminate the need for human involvement:
\textbf{(i)} it automatically identifies callbacks by simultaneously analyzing both the \android framework and the app under test;
\textbf{(ii)} it uses a combination of under-constrained symbolic execution (\textit{primitive} arguments), and \textit{type-guided} dynamic heap introspection (\textit{object} arguments) to generate valid and effective inputs.
Lastly, \tool integrates two novel feedback mechanisms---\feedbackdatadependency and \feedbackcrashguidance---during testing to increase the likelihood of triggering crashes and maximizing coverage.
In our evaluation, \tool outperforms state-of-the-art model-driven, checkpoint-based, and callback-driven testing tools both in terms of crashes and coverage.
\end{abstract}
\section{Introduction}
\label{sec:intro}

\android is the most popular mobile operating system, with $2.8$B active users and a global market share of $75\%$ as of $2021$~\cite{android_stats}.
\android apps cater to diverse users' needs, such as emailing, banking, gaming, \etc{}
The Google Play Store, the official \android app market, witnessed enormous growth---it currently hosts $2.9$M apps, and more than $100$K apps are added every month~\cite{android_app_release_stats}.
In order to provide a smooth user experience, these apps need to be thoroughly tested before developers push them to the market.
Modern \android apps use rich user interface (UI) and complex app logic, 
thus making automated exploration challenging.

\android apps are event-driven programs, \ie, each interaction with the UI of the app generates an event, which drives the app through different states.
Therefore, synthesizing a correct sequence of events is essential to efficiently explore the state space of an app.
Many prior techniques rely on UI testing frameworks~\cite{monkey,uiautomator,Baek2016,stoat,Wontae2013,Yang2013,ape} to exercise the app by generating appropriate events.
However, a large class of events is widget-specific, and requires multiple user actions to be taken in a specific order at specific UI coordinates.
As we explain in \Sec{sec:motivation}, the \code{onDateChanged} event of the \code{DatePickerDialog} widget is one such example.
Generating such events \textit{deterministically} is challenging for a UI-based testing tool, unless it has been equipped with the knowledge of how to generate all the correct events.
Given the variety of the \android widgets, and the different types of events they support, this is non-trivial.
To address this, callback-driven approaches~\cite{ehbdroid} leverage the fact that when a UI event is triggered, the associated \textit{event handler}, also known as \textit{callback}, is executed.
Callbacks are the methods in the app typically invoked by the \android framework on the occurrence of an event, \eg, \code{click} on a widget.
Callback-driven techniques call those callbacks directly---essentially eliminating the need for event generation altogether.
\looseness=-1

Existing callback-driven approaches suffer from two main limitations.
\textbf{(L1)}
They assume the knowledge of both the \android callbacks and the APIs to determine \textit{what} to call and \textit{how}, respectively.
Given an app, the first challenge is to identify its callbacks.
For that, existing tools maintain a fixed and often small list of supported callbacks.
Once a callback is identified, it has to be invoked with arguments that match the types that the callback expects.
Callbacks accept two types of arguments: \textit{primitive}, \eg, \code{int}, and \code{float}, or \textit{objects}.
Object arguments are harder to deal with.
Prior techniques depend on a human expert for writing the necessary \textit{driver} code, which would leverage widget-specific \android APIs to retrieve live objects from the app context, so that those can be supplied as arguments.
Since adding support for a callback requires a non-trivial manual effort, it is hard to extend the support for all the callbacks in the framework.
Quite understandably, while there are approximately $19,647$ callbacks in \android $4.2$~\cite{edgeminer}, the state-of-the-art callback-driven testing tool \ehbdroid~\cite{ehbdroid} supports only $58$ of them.
\textbf{(L2)} Apps accept user-supplied data as input, \eg, text.
Only generating event sequences, which existing tools focus on, is not enough, 
because certain functionalities may only be reachable under specific input.
For example, a payroll app calculates tax differently depending on the income of an employee.

This paper presents \tool, an \android app testing technique that addresses both the challenges.
To address \textbf{L1}, \tool adopts a two-phase approach.
First, we statically identify all the callbacks present in the app (\textit{what} to call).
Specifically, our \modulecallbackdiscovery module statically extracts all the callback signatures $\mathcal{L}$ supported by the \android framework.
Since an app has to override a framework callback 
to provide its own implementation, we use $\mathcal{L}$ to identify the callback implementations present in the app.
Once callbacks are identified, then we dynamically prepare arguments (\textit{how} to call) to invoke them with.
Unlike previous techniques that rely on manually-written, callback-specific driver code to generate object arguments, we resort to a hybrid approach.
Our \moduleexploration module performs a dynamic introspection of the app's heap at run-time, followed by a \textit{type-guided} object filtering to supply appropriate arguments to the callback.
This callback discovery and argument generation strategies together insulate \tool from the complexity of the \android API and obviate the need for any prior knowledge.
To address \textbf{L2}, we leverage the fact that many user inputs are of primitive types, and often appear as the arguments to the callbacks.
Therefore, the \moduleargumentgeneration module symbolizes the primitive arguments of a callback, and performs an under-constrained symbolic execution to generate the possible values of those arguments to drive the execution along all paths.
Symbolic execution is scoped within a single callback instead of the entire app to maintain a balance between precision and scalability.
\looseness=-1

In addition to tackling those two limitations, we integrate two novel feedback mechanisms into our exploration loop.
\textbf{(i)} The \modulecallbackdependency module passes on statically-identified data-dependencies between callbacks as feedback, which enables \tool to generate callback sequences that increase the likelihood of triggering crashes due to uninitialized objects, \eg, \npe.
\textbf{(ii)} We design a \textit{crash-guided} dynamic scoring mechanism that gradually deprioritizes crash-inducing paths in the app to drive the exploration towards unexplored code.
In effect, \tool is incentivized to discover more crashes than rediscovering the already found ones.
\looseness=-1

We evaluated \tool on $\numandrotestappsworked$ apps of the \androtest~\cite{androtest} benchmark, and top $\numrealworldapps$ real-world apps from the Google Play Store.
Compared to the state-of-the-art model-based techniques \stoat~\cite{stoat} and \ape~\cite{ape}, checkpoint-based technique \timemachine~\cite{timemachine}, and callback-driven technique \ehbdroid~\cite{ehbdroid}, \tool achieves $\FPeval{\v}{round(\columbusaveragecoverage-\stoataveragecoverage, 0)}\v\%$, $\FPeval{\v}{round(\columbusaveragecoverage-\apeaveragecoverage, 0)}\v\%$, $\FPeval{\v}{round(\columbusaveragecoverage-\timemachineaveragecoverage, 0)}\v\%$, and $\FPeval{\v}{round(\columbusaveragecoverage-\ehbdroidaveragecoverage, 0)}\v\%$ more in average coverage, and discovers $\FPeval{\v}{round(\numcrashescolumbus/\numcrashesstoat, 2)}\v$, $\FPeval{\v}{round(\numcrashescolumbus/\numcrashesape, 2)}\v$, $\FPeval{\v}{round(\numcrashescolumbus/\numcrashestimemachine, 2)}\v$, and $\FPeval{\v}{round(\numcrashescolumbus/\numcrashesehbdroid, 2)}\v$ times more  crashes on the \androtest apps, respectively.
\tool is also able to find $\num{\totalrealworldcrash}$ crashes in $\numrealworldappswithcrash$ real-world apps.
\looseness=-1

\medskip 
\noindent In summary, this paper makes the following contributions:

\vspace{0.7mm}
\mypar{Callback exploration.}
We propose a callback-driven \android app testing approach by presenting \textbf{(i)} a generic technique to extract all the callbacks present in an app (\Sec{sec:callback-iden}), and \textbf{(ii)} an analysis based on under-constrained symbolic execution (primitive arguments) (\Sec{sec:arg-gen}), and \textit{type-guided} dynamic object filtering for generating valid arguments to invoke callbacks.

\vspace{0.7mm}
\mypar{Feedback mechanism.}
Further, we make the app exploration systematic by integrating two novel feedback mechanisms: \textbf{(i)} a \feedbackdatadependency feedback that increases the probability of triggering bugs (\Sec{sec:callback-dep}) due to uninitialized variables, and \textbf{(ii)} a \textit{crash-guided} dynamic scoring mechanism that prevents us from rediscovering the same bugs (\Sec{sec:exploration}).

\vspace{0.9mm}
\mypar{Tool \& evaluation.}
We implement the proposed technique in a practical tool called \tool, and we make it publicly available~\cite{columbus-code}.
Our evaluation demonstrates that \tool outperforms the state-of-the-art tools both in terms of code coverage and the number of unique crashes that it identifies (\Sec{sec:eval}).
\looseness=-1

\section{Background}
\label{sec:background}

\mypar{Android events.}
\android apps are event-driven programs.
That is, apps behave as state machines, and events cause a transition from one state to the other.
An event is generated in response to one or more user actions (UI events), or by \android itself (system events).
Examples of UI events include \code{click}, \code{drag}, \code{pan}, \code{pinch}, \code{zoom}, \etc{}
Modern \android devices are equipped with peripherals, such as, Bluetooth and WiFi, and sensors like motion sensors and accelerometers.
Any change in the state of these devices is detected by the OS, which then generates a system event to notify ``interested'' apps.
Examples of system events are Bluetooth disconnected, phone tilted, and low battery level.

Based on the number of actions needed to generate an event, we define two types of events: \textit{primitive} and \textit{composite}.
Primitive events are either system events or UI events generated due to a  single action. For example, \code{MotionEvent} (\code{ME}) reports the movement of an input device like a mouse, pen, finger, trackball, or \code{KeyEvent} reports key and button related actions.
A composite event consists of multiple primitive ones, which are sequenced with strict spatial and temporal requirements.
Say, we want to drag an object from point $p_1$, and drop it at point $p_n$ along the trajectory $[p_1, p_2, p_3, ..., p_n]$.
In order to programmatically generate a \code{drag} event, the following sequence (temporal) of primitive events need to be fired at those exact coordinates (spatial): $\code{ME.ACTION\_DOWN}\; (p_1) \rightarrow \{\code{ME.ACTION\_MOVE}\; (p_i) \;|\; 2 \leq i \leq (n-1)\} \rightarrow \code{ME.ACTION\_UP}\; (p_n)$.
Without the support for a composite event, it is nearly impossible for a UI testing tool to generate most of them just `by chance'.
To make matter worse, numerous such composite events are widget-specific, \eg, the \code{DateChanged} event recognized by \code{DatePickerDialog}.
Therefore, adding support for individual events in a UI testing tool is nearly impossible.

\mypar{Android callbacks.}
An \android \textit{callback}, also known as an \textit{event handler}, is a piece of code that the framework invokes when a specific event takes place, for example; the \code{onClick} callback is called when a \code{click} event occurs.
Typically, the framework only provides empty callbacks, which an app selectively overrides to respond to the respective events.
When an event is generated, it is broken down into \code{Messages}, which are then put into a \code{MessageQueue} managed by the \code{Looper}, the entity that runs the message loop.
The \code{Looper} processes the \code{Messages} in first-in-first-out order, and calls the associated callbacks.
While invoking a callback, the framework supplies the appropriate arguments, which can be of two types---\textit{primitive}, \eg, \code{int}, \code{float}, \etc, or \code{object}, \ie, an instance of a class.
\looseness=-1

\mypar{Android \activity}:
An \activity is a UI element that acts as a container of other UI elements.
It often presents itself in the form of a window.
Activities are managed by maintaining an activity stack.
When a new \activity starts, it is placed on the top of the stack,
while the previous one is paused, and remains below the current one in the stack.
A paused \activity does not come to the foreground again until the current \activity exits.
An \activity transitions through different states of its \textit{lifecycle} as a user navigates through an app.
Lifecycle callbacks, \eg, \code{onCreate}, \code{onPause}, \code{onResume}, are the ones associated with such lifecycle events.
 
\section{Motivation and challenges}
\label{sec:motivation}
This section introduces a motivating example, the challenges it presents to the state-of-the-art callback-driven app testing tools, 
and how we tackle them.

\begin{figure}[t]
	\begin{minipage}{.4\columnwidth}
		\vspace{1mm}
		\includegraphics[width=0.8\linewidth]{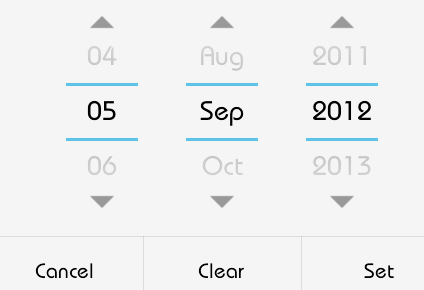}
		\captionsetup{justification = raggedright, singlelinecheck = false}
		\vspace{3mm}
		\captionof{figure}{A \code{DatePickerDialog} widget}
		\label{fig:datepicker}
	\end{minipage}%
	\begin{minipage}{.55\columnwidth}
\begin{lstlisting}[
			language=java,
			frame=single,
			xleftmargin=10pt,
			caption=,
			stepnumber=1,
			firstnumber=1,
			basicstyle=\ttfamily\tiny,
			linebackgroundcolor={%
				\ifnum\value{lstnumber}=2
				\color{orange!25}
				\fi
				\ifnum\value{lstnumber}=7
				\color{orange!25}
				\fi
			}
			]
void onCreate(Bundle bundle) {
  ListView lv = getListView();
}

void ehbTest() {
  for (int i=0; i<lv.size(); i++) {
    View v = lv.getChildAt(i);
    long id = lv.getAdapter()
                .getItemId(i);
    this.onListItemClick(lv,v,i,id);
  }
}
\end{lstlisting}
			\vspace{-4mm}
			\captionof{figure}{\ehbdroid instrumentation for \code{onListItemClick()}}
			\label{fig:instrumentation}
	\end{minipage}
\vspace{-3mm}
\end{figure}

The code in \Fig{fig:motivation} shows three callbacks that an \android app might implement. The callback functions are executed when the user interacts with specific UI elements, \ie, clicks on a list item, clicks on a button, and sets a date using a \code{DatePickerDialog} (\Fig{fig:datepicker}), respectively.
UI-based testing tools~\cite{monkey} generate events, \eg, clicks, to interact with the UI of such apps.
However, these tools are not widget-aware, meaning that, they are unable to \textit{systematically} generate composite events unless they already know how to generate them.
For example, the following events need to be generated in an exact sequence, on specific UI elements, to call the \code{onDateChanged} callback---\textbf{(i)} \code{DatePickerDialog} widget is clicked to bring up the spinner control, 
\textbf{(ii)} the day/month/year is changed by clicking on the up/down arrows, and
\textbf{(iii)} the \code{Set} button is clicked.
It is unlikely for a UI-based testing tool to be able to deterministically generate this event sequence without any guidance.
Moreover, to set a particular date, the up/down arrows need to be clicked a specific number of times---which is hard as well.
To overcome this limitation, callback-driven techniques~\cite{ehbdroid} invokes the callback, \eg, \code{onDateChanged}, directly bypassing the UI layer altogether.
While callback-driven testing shows promise, it still suffers from the following limitations.

\mypar{Identifying callbacks.}
The first step of callback-driven testing is identifying the callbacks.
Unfortunately, the set of callbacks supported by the \android framework is huge.
While previous research~\cite{edgeminer} identified approximately $19,647$ callbacks in \android $4.2$; \ehbdroid, the state-of-the-art callback-driven testing tool, supports only $58$ callbacks.
\tool statically analyzes the app and the \android framework together to address this issue (\Sec{sec:callback-iden}).

\begin{figure}[t]
\begin{lstlisting}[
	language=java,
	xleftmargin=10pt,
	frame=single,
	caption=,
	stepnumber=1,
	firstnumber=1,
	basicstyle=\ttfamily \scriptsize,
	linebackgroundcolor={%
		\ifnum\value{lstnumber}=4
		\color{blue!25}
		\fi
		\ifnum\value{lstnumber}=10
		\color{blue!25}
		\fi
		\ifnum\value{lstnumber}=18
		\color{orange!25}
		\fi
	}]
protected void onListItemClick(ListView l, View v, int position, long id) {
  File f = (File)(mList.get(id).get(ITEM_KEY_FILE));
  if (f.isFile()) {
    mSelectedFile = f;
    showDialog(DIALOG_IMPORT_FILE);
  } 
}

public void onClick(DialogInterface dialog, int whichButton) {
  File f = mSelectedFile;
  Intent i = new Intent(mContext, myActivity.class);
  Uri u = Uri.fromFile(f);
  i.setData(u);
  startActivity(i);
}

public void onDateChanged(DatePicker view, int year, int month, int day) {
  if (day == 15 && month == 6 && year == 2020)
    Toast.makeText(context, "Success!", ...).show();
}
\end{lstlisting}
	\vspace{-4mm}
	\caption{\small Code containing three callbacks.
		Their data dependencies (\crule[blue!25]{2mm}{2mm}) and checks on the arguments (\crule[orange!25]{2mm}{2mm}) are highlighted.}
	\label{fig:motivation}
	\vspace{-2mm}
\end{figure}

\mypar{Providing callback arguments.}
Callbacks accept either primitive arguments or objects.
The primitive arguments are often involved in path conditions within the callback.
Without the correct value of such primitives, part of the callback may never be exercised.
In \Fig{fig:motivation}, the \code{Toast} message appears only on a specific date.
Existing callback-based testing tools use a set of predefined values to invoke callbacks.
Therefore, \Ln{19} will possibly never be explored.
\tool symbolizes primitive arguments and employs under-constrained symbolic execution to infer values to make larger part of the callback code reachable (\Sec{sec:arg-gen}).

For object arguments, such as, the \code{ListView} and \code{View} arguments of the \code{onListItemClick} callback in \Fig{fig:motivation}, callback-driven tools use the \android API (by statically instrumenting the app) to retrieve correct objects from the app context, as shown in \Fig{fig:instrumentation} (\Ln{2} and \Ln{7}).
However, this approach is not scalable, as the number of callbacks in the \android framework is huge, and the tool requires adding explicit support for all the arguments of all the callbacks.
Instead, \tool retrieves live objects from the app heap at runtime, and then applies \textit{type-guided} object filtering to provide the correct arguments (\Sec{sec:arg-gen}).
Type information comes from a one-time, static, pre-processing phase.

\mypar{Data dependency feedback.}
Variables are often shared among multiple callbacks.
Shared data introduces data dependencies, which an app should either enforce by restricting available UI actions, or handle by placing a sanity check.
In \Fig{fig:motivation}, both the \code{onClick} and \code{onListItemClick} callbacks use the same variable \code{mSelectedFile}.
Specifically, \code{onListItemClick} opens a file, and sets the file handle \code{mSelectedFile} (\Ln{4}), which \code{onClick} uses in \Ln{10}.
This implies that \code{onListItemClick} has to be invoked before \code{onClick}, otherwise the \code{onClick} method would generate a \npe.
\tool statically infers such data dependencies and passes the same as feedback during testing.
While synthesizing a callback sequence, \tool attempts to violate the expected order to increase the likelihood of inducing crashes (\Sec{sec:callback-dep}).

\section{The \tool{} Framework}
\label{sec:framework}

\begin{figure}[t]
	\centering
	\includegraphics[width=.8\columnwidth]{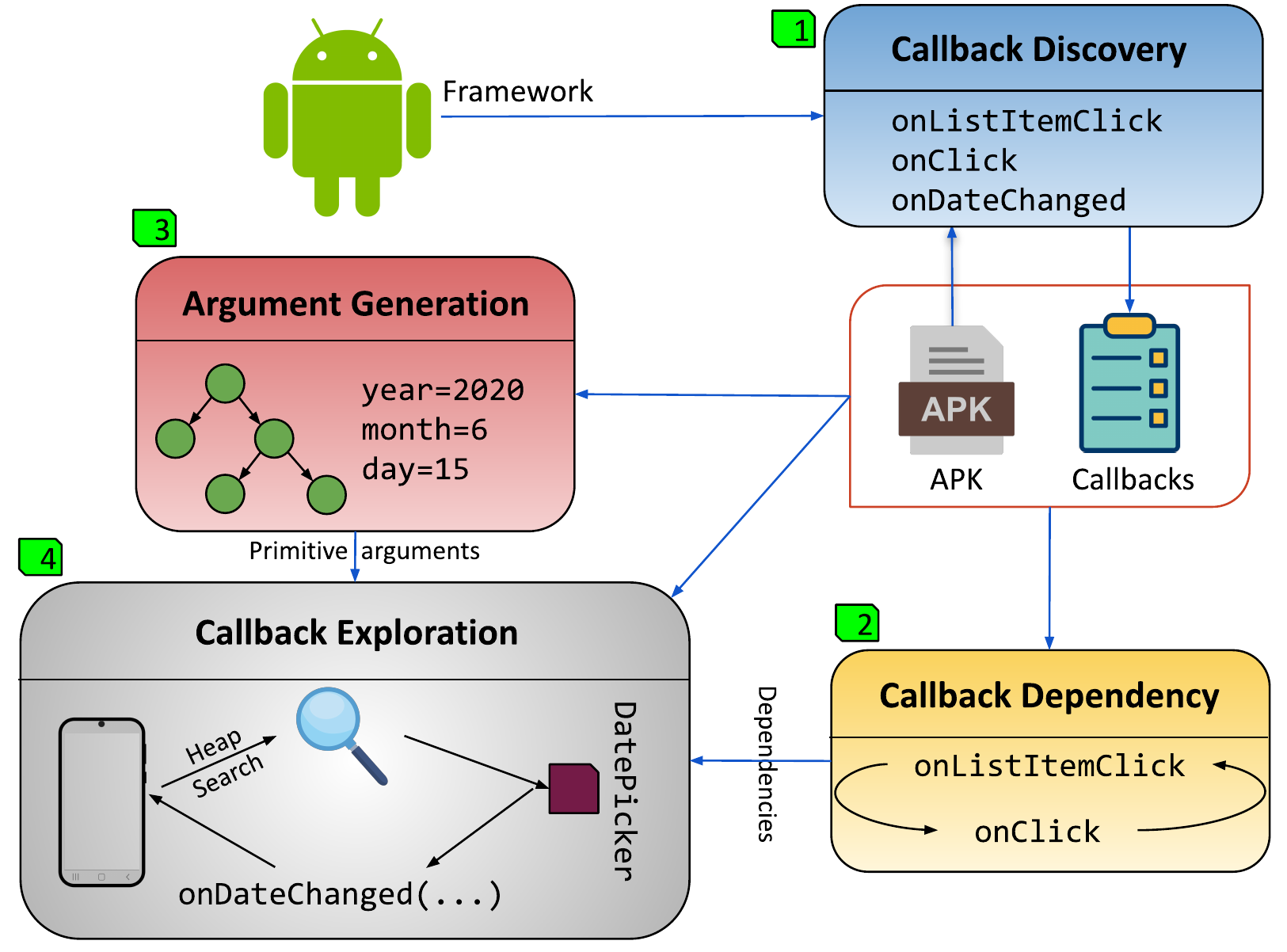}
	\vspace{-2mm}
	\caption{Overview of \tool with reference to the motivating example in \Fig{fig:motivation}}
	\label{fig:approach}
	\vspace{1mm}
\end{figure}

In this work, we propose \tool, a framework to test \android apps by directly invoking their callbacks.
For a given \android app, \tool first identifies its callbacks (\Sec{sec:callback-iden}).
It then obtains the primitive argument values that correspond to different execution paths in these callbacks (\Sec{sec:arg-gen}) and identifies inter-callback dependencies (\Sec{sec:callback-dep}).
Finally, our tool invokes the identified callbacks---\textbf{(i)} in orders that initially violate (to increase the chances of triggering uninitialized data-related bugs), and later respect their dependencies, \textbf{(ii)} with their expected arguments during the exploration (\Sec{sec:exploration}).
\tool keeps track of the callback-defining classes explored during the app execution, and gives higher priority to exploring classes that have been less explored. \Fig{fig:approach} depicts the high-level workflow of our system.

\subsection{Callback discovery}
\label{sec:callback-iden}

Every \android app defines its own set of callbacks.
Though state-of-the-art approaches~\cite{ehbdroid} resorted to a predefined set of callbacks, the \android framework contains thousands~\cite{edgeminer} of callbacks, and the number is constantly increasing.
In order to facilitate effective app exploration, in this work, we present an  approach to automated callback discovery.
\tool's callback identification is presented in \Alg{alg:callback}. 
At a high level, our callback discovery approach first statically analyzes the framework (Function \fanalysis) followed by an analysis of the app under test (Function \aanalysis), and outputs a list of callbacks present in the app.

\setlength{\textfloatsep}{0.5pt}
\begin{algorithm}
\scriptsize
\caption{Static callback identification}
\label{alg:callback}
\Fn \fanalysis {
\Input{Android framework JAR}
\Output{Classes with callback candidates $\Delta$}
$\Delta \gets \{\}$\;
$CG_f \gets \getcg(\textup{JAR})$\;
$CH_f \gets \getch(\textup{JAR})$\;
\ForEach{ \textup{class} $c_f \in \getclassesj(\textup{JAR})$ } {
$M_f \gets \emptyset$\;
\ForEach{ \textup{method} $m_f \in \getmethodsC(c_f)$ } {
    \If{$\getmethodacc(m_f)$} {
        \If {$\getcallers(c_f, m_f, CG_f) \neq \emptyset$ } {
            $M_f \gets M_f \cup m_f$
        }
    }
}
 $\Delta[c_f] \gets \Delta[c_f] \cup M_f$\;
}
 \ForEach{ $(c_f, M_f) \in \Delta$ } {
    \ForEach{ \textup{subclass}  $c_f^\prime \in \getsubclasses(c_f)$ } {
        $M_f^\prime \gets \Delta[c_f^\prime]$; 
        $M_f^\prime \gets M_f^\prime \cup M_f$; 
        $\Delta[c_f^\prime] \gets M_f^\prime$\;
    }
 }
 \KwRet $\Delta, CH_f$
}
\Fn \aanalysis {
\Input{App's APK, Framework classes with callback candidates $\Delta$, Framework's class hierarchy $CH_f$}
\Output{Application callbacks $CB$}
$CB \gets \emptyset$\;
\ForEach{ \textup{class} $c_a \in \getclassesa(\textup{APK})$ } {
    $ClassAndItsParents \gets c_a \cup \getsuperclasses(c_a)$\;
    \ForEach{ $cp_a \in ClassAndItsParents$ } {
        \ForEach{ $(c_f, M_f) \in \Delta$ } {
            \If{$cp_a \kwextends c_f \lor cp_a \kwimplements c_f$} {
                \ForEach{$m_a \in \getclassmethods(cp_a)$}{
                	\ForEach{$m_f \in M_f$}{
		                \If{ $\iscompatible(m_f, m_a)$ }{
		                     $CB \gets CB \cup m_a$\;
		                }
	                }    
                }
            }
        }
    }
}
 \KwRet $CB$
}
\end{algorithm}

\mypar{\android framework analysis.}
Our analysis is based on two observations.
\textbf{(i)} As discussed in \Sec{sec:background}, in order to perform the intended action once an event is generated, an app needs to override the respective callback present in the \android framework.
To be overridden, a callback needs to be declared as either a \textit{protected}, or a \textit{public} method within the framework.
\textbf{(ii)} Moreover, at runtime, callbacks are typically invoked within the framework through a series of internal method calls once an event is generated---meaning that, callbacks have caller(s) within the framework.

\tool first constructs the framework's callgraph $CG_f$.
To build the call graph, \tool performs intra-procedural type inference~\cite{Palsberg1991ObjectorientedTI} to determine the possible dynamic types of the object on which a method is called.
When this fails, \tool then over-approximates the possible targets as all the subclasses of its static type.
Now, for every method $m_f$ in a framework class $c_f$, \tool considers $m_f$ as a potential callback (\Lns{7}{13}) if---\textbf{(i)} $m_f$ is declared as either \textit{protected}, or \textit{public}, and \textbf{(ii)} $m_f$ has at least one caller in $CG_f$.
At the end, we compute a mapping $\Delta$ that maps each class $c_f$ to their potential callbacks.
Each callback $m_f$ is a tuple, which consists of the defining class $c_f$, the method name, and the types of its arguments.
Now, this mapping $\Delta$ is incomplete, because a class can inherit callbacks from its superclasses as well.
Therefore, \tool computes the complete list of potential callbacks for every $c_f$ by walking up the class hierarchy to consolidate superclass callbacks, too (\Lns{16}{20}).
The updated callback mapping $\Delta$ and the class hierarchy information $CH_f$ are returned as the output.
Note that \tool performs the framework analysis once per framework.
\looseness=-1

The above analysis is inspired by EdgeMiner~\cite{edgeminer}.
The main goal of EdgeMiner is to detect framework callbacks, and using that to discover the registration methods within the framework.
However, the end goal of Columbus is to detect application level callbacks by leveraging the framework callbacks.

\mypar{\android app analysis.}
The goal of this phase is to find whether any app class method $m_a$ is a valid overriding method of the framework class callback $m_f$.
In order to override a callback within an app, the app class $c_a$ needs to either extend or implement the corresponding callback-defining class $c_f$ of the \android framework.
For example, in \Fig{fig:motivation}, to override the \texttt{onListItemClick} callback, the app class needs to extend the \texttt{ListActivity} framework class.
\tool identifies such pairs of classes $(c_f, c_a)$ by statically analyzing the app.
In the next step, it checks whether any app method $m_a \in c_a$ has the same name and the same number of arguments as any framework method $m_f \in c_f$, and the arguments of $m_a$ are \textit{type-compatible} with those of $m_f$ (\Lns{29}{35}).
We call a type $t_1$ to be \textit{compatible} with another type $t_2$, if either $t_1 = t_2$, or $t_1$ is a subclass of $t_2$ according to the class hierarchy.
To determine type compatibility, \tool constructs the full class hierarchy by unifying ($\oplus$) the framework class hierarchy $CH_f$ with the app class hierarchy $CH_a$.
Let $A \rightarrow B$ denote that $A$ is a superclass of $B$.
Now, if the relations $H_1 = A \rightarrow B$ and $H_2 = B \rightarrow C$ appear in $CH_a$ and $CH_f$, respectively, then $H_1 \oplus H_2 = A \rightarrow B \rightarrow C$.
Finally, we obtain the set of potential callbacks in an app.
Our analysis would discover all three functions \code{onListItemClick}, \code{onClick}, and \code{onDateChanged} in \Fig{fig:motivation} as callbacks.

Identifying callbacks by analyzing either the app, or the framework alone is challenging.
Since a callback is invoked by the framework, the callback methods do not have incoming edges visible from the call graph of the app.
However, an analysis relying only on this fact alone will generate false positives---because, it could detect a non-callback method as a callback due to the inherent incompleteness of Java call graphs~\cite{java_call_graph}.
Similarly, our framework analysis is over-approximated in a way that will  definitely contain the callbacks, but non-callbacks methods, too.
Intuitively, therefore we `intersect' the framework callback candidates and app methods to determine the true callbacks.

During this phase, we can encounter methods of a generic Android framework class \texttt{Object}, that are declared as \texttt{public}, and can therefore be overridden by the corresponding application-level classes inheriting the \texttt{Object} class.
The number of such callbacks appearing as part of the final callback list was negligible (around 3\%).
We do not consider such methods as callbacks.

\subsection{Generating arguments for callbacks}
\label{sec:arg-gen}
In order to invoke a callback, we need to provide argument values conforming to the correct types.
In case of GUI-action-driven exploration strategies, the framework provides these arguments, which are derived from the events resulting from the GUI actions.
Therefore, to invoke callbacks {\it without} relying on GUI actions, \tool needs to tackle the challenge of generating arguments for these callbacks, with a goal  to explore the paths within a callback resulting in faster coverage and better crash discovery.

A callback argument can be one of two types: primitive or reference.
For each type, \tool uses different strategies to generate the corresponding arguments.

\subsubsection{Primitive type arguments.}
Primitive type arguments, \eg, \code{integer}, \code{long}, \code{string}, and \code{boolean}, are typically involved in program paths that can only be explored with a specific set of values.
For instance, \Ln{19} of the \code{onDateChange} callback in \Fig{fig:motivation}  will get executed \textit{only if} the integer arguments \code{day}, \code{month}, and \code{year} are equal to $15$, $6$, and $2020$.
Therefore, to effectively explore all the paths in such a callback without resorting to a computationally expensive random search,
\tool needs to provide these specific set of values to the callback during invocation.
In this case, \tool symbolizes respective callback arguments, and performs an under-constrained symbolic execution (until termination, or time-out) to generate concrete values.
\looseness=-1

Precisely, \tool starts the symbolic execution at the entry point of each of the callbacks, and collects constraints on the arguments corresponding to each of the execution paths.
It then solves these constraints and generates concrete argument values, which when provided as arguments to the callback during invocation, result in exercising those paths within the callback.
During symbolic execution, we track constraints on objects that modify the program state, such as \textbf{(i)} callback arguments, and \textbf{(ii)} API return values.

\mypar{Callback arguments.}
\tool executes the callback with symbolic and unconstrained arguments.
It then collects the constraints in each of the execution paths that involve operations on the symbolic arguments.
For example, if one of the arguments is an object, and during execution, one of its fields is set to $5$, \tool's symbolic execution engine will automatically add a  constraint stating that the specific attribute needs to be equal to $5$ (to follow a particular program path of interest).

\mypar{API calls.}
\tool's symbolic execution engine generates summaries for common functions, for example, the Java runtime function \code{exit()}.
These summaries capture the side effects of these APIs that modify the program state.
For APIs without a summary, we return a fresh symbolic value conforming to the return type of the API.

\tool's symbolic execution engine is capable of generating concrete values of  \code{integer}, \code{float}, \code{boolean}, and constant \code{string} types.

\subsubsection{Reference type arguments}
Reference type argument objects frequently represent UI elements where a user performs certain actions.
In \Fig{fig:motivation}, when a user clicks on \texttt{AlertDialog} (a subclass object of \code{DialogInterface}), the framework invokes the \code{onClick} callback with an argument object of type \texttt{AlertDialog}.
Therefore, to invoke the \code{onClick} callback without relying on the \android framework, we need to provide an object of type \code{DialogInterface}, or a subclass of \code{DialogInterface}---as an argument.

\mypar{App heap search}.
During the app exploration (\Sec{sec:exploration}), as and when new \texttt{Activities} are visited, these object instances are created in the app heap.
Therefore, in order to invoke a callback that requires reference type arguments, \tool monitors the app heap by dynamically instrumenting the app under test.
In many cases, the argument type present in the callback signature is not the one created in the app heap.
In \Fig{fig:motivation}, the \code{onClick} callback has an argument of type \code{DialogInterface}.
However, the object created will be of type \code{AlertDialog}, a subclass of \code{DialogInterface}.
To account for this scenario, \ie, if an object instance of a reference type inferred from the callback signature is not available in the app heap, \tool searches for object instance(s) that is a subclass of the required type.

\mypar{Custom object creation.}
It may still happen that no object instances of the required type or its subclass are found in the heap.
For example, certain types of objects required as a callback argument, \eg, \code{KeyEvent}, and \code{MotionEvent}, that are created by the \android framework \textit{only} when it registers touch, or key-press on UI elements.
Therefore, in order to invoke such callbacks, \tool leverages Java reflection.
Specifically, for such a reference, \tool creates the object using its public constructor.
If the constructor expects primitive type arguments, \tool uses either a random value, or a value from a pre-defined set as the argument.
For example, to create \code{KeyEvent}, or \code{MotionEvent} objects, \tool uses pre-defined values as they should be valid screen coordinates in order to successfully explore the callback.
If a constructor expects reference type objects, \tool either finds these objects through app heap search, or creates recursively through Java reflection.
For example, if we were to create an object of type \code{A} which has a constructor that accepts an object of type \code{B}, then we create objects bottom up (\ie, first \code{B}, then \code{A}).
In case multiple such constructors exist, \tool picks the one which requires the least number of reference type arguments.

\noindent
\subsection{Inter-callback dependency}
\label{sec:callback-dep}
\noindent
Callbacks within an app can share variables resulting in \textit{read-write} data dependencies.
As discussed in \Sec{sec:motivation}, for \code{onListItemClick} and \code{onClick} callbacks (\Fig{fig:motivation}), prioritizing dependency-violating order, \ie, invoking \code{onClick} before \code{onListItemClick}, brings us faster to a crash discovery.
Whereas invoking the callbacks in the dependency-respecting order allows for a better code coverage. For example, the execution of the \Lns{13}{14} in \code{onClick} happens only if the reference \code{mSelectedFile} accessed at \Ln{10} is defined by a prior execution of \code{onListItemClick}.

Based on this observation, \tool computes callback pairs having shared variable dependencies by performing a field-insensitive analysis of the app. %
The intuition is to first compute a set of class variables $vars$ that are \textit{not} initialized through a \textit{default initializer}.
The \textit{default initializer}s are the methods that get automatically invoked whenever a class or activity gets created, \eg, the life cycle methods of an activity, class constructors, \etc{}
These variables $vars$ are our target candidates, since they are defined and accessed only through callbacks.
Next, for every such variable $var \in vars$, \tool searches for callback pairs $(cb_1, cb_2)$ where one of them \textit{read}s (\code{R}) $var$, and the other \textit{write}s (\code{W}) $var$.
The output of this phase will be a set of variables with their dependent callback pairs.
For the example in \Fig{fig:motivation}, the output will be $\{\code{mSelectedFile}, (\code{`R'}, \code{onClick}), (\code{`W'}, \linebreak[1]\code{onListItemClick})\}$.

These dependency pairs are used as feedback during the exploration phase detailed in \Sec{sec:exploration}.
In order to accelerate crash discovery, \tool implements a weighted-score based exploration strategy, which initially prioritizes executing callbacks that write to variables over the callbacks that read from the same variables---inducing the dependency violating callback invocation orders.
However, during the exploration, \tool dynamically adjusts the scores, \eg, penalizes the callbacks that frequently result in a crash, or prioritizes the callbacks that are executed less frequently, in order to explore newer or less explored program paths as well.

\subsection{Callback-guided exploration}
\label{sec:exploration}

To explore an app under test, we first statically obtain its callbacks (\Sec{sec:callback-iden}), their dependencies (\Sec{sec:callback-dep}), and the primitive argument values (\Sec{sec:arg-gen}).
Then, \tool spawns the app, dynamically instruments it to inspect the app heap, and starts exploring its functionalities.
\tool invokes a callback whenever an instance of the \activity, or the class defining the callback appears in the app's heap.
If the callback expects reference type arguments, \tool then generates such argument objects using the strategy detailed in \Sec{sec:arg-gen}.
\Alg{alg:exploration} gives an overview of our app exploration strategy.
\tool's exploration strategy is composed of the following components:

\mypar{Activity monitor.}
As the app is being explored, two kinds of entities get created, or destroyed in the heap: 
\textbf{(i)} activities and related UI element objects, and
\textbf{(ii)} regular class objects, as the side-effect of calling a callback that instantiates the class.
The activity monitor records such events by monitoring the invocation of the lifecycle callbacks of the activities, and the class constructors.
For example, invocation of \code{onCreate()} signals an activity creation, and \code{onDestroy()} is invoked when an activity is destroyed.
The activity monitor maintains an activity stack $\mathcal{S}$ by pushing an activity to $\mathcal{S}$ when a new activity is created, and popping an activity off $\mathcal{S}$ when it is destroyed.
Therefore, the most recently created activity, which we call as the \textit{live} activity, always remains at the top of $\mathcal{S}$.

The app is explored in a depth-first manner, and runs in continuous \textit{cycles}.
For a live activity $act$, the activity monitor retrieves all the class objects $newClasses$ created in the app heap (\Ln{18}), passes it on to the \textit{selector} for choosing the next callback $cb$, which is then executed by the \textit{executor}.
The function \code{getNewClasses()} returns only those classes for which at least one callback is still unexplored.
If a callback creates a new live activity $act^\prime$, the activity monitor puts $act$ on hold, and switches to $act^\prime$.
When all the callbacks of an activity or its associated classes have been executed, the activity monitor destroys the activity, removes it from $\mathcal{S}$ (\Lns{19}{22}), and starts exploring the next live activity.
One testing cycle ends, and the next one begins when $\mathcal{S}$ becomes empty.

\mypar{Selector.}
The selector module receives the candidate classes $newClasses$ to be explored from the activity monitor, and chooses a callback $cb$ to be executed next (\Ln{24}).
While choosing $cb$, it considers the class weights $ClW$, callback weights $CbW$, inter-callback dependencies $Dep$, and the visited status $explored$ of the callbacks.
The $explored$ map is cleared when a testing cycle begins.
All the weights are initially set to zero, and are dynamically adjusted during the exploration based on how frequently the classes and the callbacks have been explored.
Similarly, when a callback is explored, the $explored$ map is updated (\Ln{26}).

To choose a callback, the selector employs multiple strategies in the following order:
\textbf{(i)} In the beginning, when none of the callback is explored, the selector uses $Dep$ to choose the callback $cb$ with the read (\code{R}) dependency, and its defining class $cl$.
\textbf{(ii)} The selector consults the $explored$ map to prioritize unexplored callbacks over the explored ones.
\textbf{(iii)} A class or callback with lower weight ($ClW$ or $CbW$) has been explored the least; therefore it is prioritized next for execution.
The tie among multiple unexplored classes, or callbacks with the same weight is broken randomly.
\begin{algorithm}
\scriptsize
\caption{Callback driven exploration}
\label{alg:exploration}

\Fn \fuzzexplore {
\Input{Application callbacks $AC$, their dependencies $Dep$, class hierarchical information $CH_f$ and $CH_a$, duration $t$}
\Output{Crash dumps $crashes$}
$crashes \gets \emptyset$, $explored \gets \{\}$, $testingCycle \gets 0$\;
$CbW \gets \emptyset$ $\;\;\;\;\;\;\;$ // \textup{callback weights}\; 
$ClW \gets \emptyset$ $\;\;\;\;\;\;\;$ // \textup{class weights}\;

\ForEach{ \textup{callback} $cb \in AC$ } {
    $cl \gets \getclassdefiningmethod(cb)$\;
    $CbW \gets CbW \cup (cl, cb, 0.0)$\;
    $ClW \gets ClW \cup (cl, 0.0)$\;
}

\While{\textup{until} $t$ \textup{is reached}}{
	$\textup{\spawnapp()}$\;
	$testingCycle \gets testingCycle + 1$ \;
	\ForEach{ \textup{callback} $cb \in AC$ } {
		$explored[cb] \gets false$\;
	}		
	
	\While{\textup{until no new activity left to explore}}{
		$act \gets \getforegroundact()$\;
		$newClasses \gets \getnewclasses(act, explored)$\;
		
		\If{$newClasses = \emptyset$}{
			$\remove(act)$\;
			$\textup{go to \Ln{16}}$\;
		}

		$cl \gets \getnextclass(newClasses \cup act, ClW, Dep)$ \\
	    $cb \gets \getnextcallback(cl, explored, CbW, Dep)$\;
	    
	    \If{$cb = \emptyset$}{
			$explored \gets explored - (cb, false) \cup (cb, true)$\;
			$\textup{go to \Ln{16}}$\;
		}
	    
	    $allargs \gets \generatearguments(cl)$ \;
	    \ForEach{$args \in allargs$ } {
	    	$inst \gets \getinstance(cl)$\;
	    	$newCrash \gets \fuzzcallback(inst, cb, args)$\;

		    \If{$newCrash \ne \emptyset$} {
		                $crashes \gets crashes \cup newCrash$\;
		                $\updatepenalizeweights(ClW,CbW,cl,cb)$\\
		                $\textup{\restartapp() and go to \Ln{10}}$\;
		            }
		    \Else{
		        $\updateweights(ClW, CbW, cl, cb)$
		    }
		}
	}
}

 \KwRet $crashes$
}
\end{algorithm}

\mypar{Executor.}
The executor executes the callback selected by the selector.
The executor searches the app heap for an instance of a class, or an \activity that overrides the callback (\Ln{31}).
If an instance is found, the executor generates the arguments for the callback respecting their types (\Sec{sec:arg-gen}).
However, an argument can have multiple possible values executing different paths (primitive), or depending on the availability of objects in the heap (reference).
The executor, therefore, schedules the callback for execution for each combination of such inferred values.
After each execution, the class weight for a class $cl$ and the callback weight for a callback $cb$ are updated as shown in \Fig{fig:update_weights}.

\begin{figure}[htbp]
	\[\begin{array}{rlll}
	CbW_{cb} & :- & CbW_{cb} + \frac{ex_{t}}{sch}\\
	& & \text{$sch$ $\gets$  number of scheduled executions of $cb$} \\
	& & \text{$ex_{t}$ $\gets$  number of executions of $cb$ at time $t$} \\
	& & \text{} \\
	ClW_{cl} & :- & avg(CW_{cb})\; \forall cb \in cl 

	\end{array}\]
		\caption{New class and callback weights after each execution}
	\label{fig:update_weights}
	\vspace{2mm}
\end{figure}

Intuitively, the executor updates the weights to reflect what percentage of callbacks are executed with respect to the total number of possible invocations---since a crash, or a creation of new activity may interrupt the processing of the rest of the scheduled executions.
The class weights are accordingly adjusted such that the least explored class, and its callbacks are prioritized to be executed the next time the activity comes live.

\mypar{Crash detector.}
After the execution of a callback, the crash detector monitors whether it results in a crash of the app.
We do not want to rediscover the same crash repeatedly.
Therefore, if a crash happens, the \code{UpdateAndPenalizeWeights()} (\Ln{35}) function updates the class weights to deprioritize the callback $cb$, and its defining class $cl$---the callback weight $CbW_{cb}$ is increased by $\delta$ (an empirically determined constant), and accordingly the class weight $ClW_{cl}$ is adjusted.
The idea is to gradually increase the callback weight in order to account for the case when \textit{only} a specific set of argument values results in a crash, and all other values should still be able to explore the callback.
Therefore, instead of not choosing the callback at all, the selector deprioritizes the callback for some time.

\section{Evaluation}
\label{sec:eval}

In our evaluation, we aim at answering the following research questions:
\textbf{RQ1.} How does \tool compare with the state-of-the-art testing tools in terms of both code coverage and discovered crashes?
\textbf{RQ2.} How effective is \tool in finding crashes in popular, real-world apps?
\textbf{RQ3.} What is the benefit of leveraging dependency feedback?

\subsection{Experimental setup}

\mypar{Dataset.}
To answer \textbf{RQ1} and \textbf{RQ3}, we used \androtest~\cite{androtest}, a collection of $\numandrotestapps$ apps.
This dataset has become the de facto standard benchmark for \android app testing, and it has been used in the evaluation of a large number of tools~\cite{androtest,stoat,sapienz,timemachine,dynodroid,trimdroid,Anand:2012,guiripper,swifthand,Baek2016,Wei13,Mahmood:2014}.
However, we had to remove $\numandrotestappsdidnotwork$ apps that were not fully compatible with \android~9 (which is the environment we used for \tool).
For example, the \code{ListView} in the \code{netcounter} app does not appear in \android 9.
Therefore, we used the remaining $\numandrotestappsworked$ apps for all our experiments.

For \textbf{RQ2}, we created a dataset of popular, real-world apps.
We will refer to this dataset as the \textit{real-world} dataset.
To build this dataset, we first compiled a list of
Google Play Store~\cite{google_play_store} apps with a minimum of $\num{\mininstalls}$ installs and a user rating of at least $\minrating$ stars.
Then, we collected first $\numrealworldapps$ apps compatible with \frida instrumentation.
As we show in \Tbl{tbl:app_types_real_world}, these apps are quite diverse and belong to $\numrealworldappcategories$ broad categories.

\mypar{Environment.}
Our experiments were conducted on a system with an Intel(R) Core(TM) i9-10885H @ 2.40GHz processor ($16$ cores), $128$GB of memory, and $1$TB of solid-state drive (relevant for the snapshot save and restore mechanism used by \timemachine), running a 64-bit Ubuntu 20.04 operating system.
For testing, we used $8$ Google Pixel 3a phones
running \android 9 (Pie, API level 28), with the Internet and Bluetooth connectivity enabled.
We did not create any accounts for those apps that allow user logins.
We ran each tool for $\testduration$ hours on each app, repeated each experiment $\numrepetitions$ times, and averaged out the results to minimize the effect of any inherent randomness.
Before testing each app, we first brought the phones to a \textit{clean-slate} state by wiping its \code{sdcard} contents, and then pushed the \code{sdcard} files used by \stoat in their experiment to the phones.
All the tools except \timemachine, which requires a virtual machine (VM) to operate, were tested on real hardware (phone).
\looseness=-1

\mypar{Pre-exploration.}
Before the dynamic exploration could begin, \tool prepares an app by running the first three static pre-processing phases.
We provide relevant results for the $\numandrotestappsworked$ apps of the \androtest dataset:
The \modulecallbackdiscovery module identified a total of $\num{\numframeworkcallbacks}$ and $\num{\numappcallbacks}$ callbacks in the \android framework and the apps, respectively.
Out of $\num{\numappcallbacks}$ app callbacks discovered, $\num{\numcallbackswithatleastoneprimitiveargument}$ callbacks had at least one primitive argument, thus necessitating the invocation of the \moduleargumentgeneration module.
With a timeout of $\symbolicexecutiontimeout$ minutes, the argument generation succeeded for $\num{\argumentgenerationsucceeded}$ callbacks, while it timed out for the remaining $\num{\argumentgenerationtimedout}$ callbacks.
Additionally, $\num{\numreferencetypecallbacks}$ callbacks have at least one reference type argument, and in total $\num{\totalreferencetypeargs}$ reference type arguments.
Out of them, $\num{\numreferenceargsonheap}$ objects were always found on the heap, and the remaining $\num{\numreferenceargscreated}$ objects needed to be created.
Finally, the \modulecallbackdependency module discovered a total of $\num{\totaldependencyrelations}$ dependency relations between $\num{\totaldependencyonvariables}$ variables across all the apps.

\mypar{Coverage and crash collection.}
We used \emma~\cite{emma} to collect statement coverage.
The coverage data was collected every minute for all tested tools.
\emma injects its own instrumentation code into the apps.
Unfortunately, its coverage reports do include coverage data from its own packages, which can either inflate, or deflate the overall coverage.
Therefore, we excluded \emma-specific classes from the coverage calculation.

We detect crashes by parsing 
\textbf{(i)} \logcat~\cite{logcat} logs fetched by the log watcher, a long-running process that streams logs from the devices (phones) in real-time, and
\textbf{(ii)} logs of the crashes captured by the \frida server.
We used the widely adopted practice of computing the stack hash to determine the \textit{uniqueness} of crashes.
Crashes that do not contain the app's package name were filtered out. 
For \frida reports, we occasionally observed that certain crashes that originate from the dynamic instrumentation contain an app's package name.
Therefore, we manually inspected and removed those irrelevant crashes after the initial package-name-based filtering.
Then, we normalized the stack traces for the remaining crashes by removing  irrelevant and ephemeral information, \eg, timestamp, process id (PID), \etc{}
Finally, we compute hashes over these sanitized stack traces.
\looseness=-1

\mypar{Implementation.}
We implemented the first three phases of our analysis, \viz, callback identification, callback dependency discovery, and primitive argument generation using the \angr~\cite{angr} binary analysis framework.
All these phases are performed offline, before the testing begins on the device.
For exploration, the final phase, we leveraged the \frida~\cite{frida} dynamic instrumentation toolkit.

\subsection{Experimental results}

\begin{table}[t]
	\tiny
	\begin{tabular}{|p{0.9cm}|p {1.7mm} p {1.7mm} p {1.7mm} p {1.6mm} p {1.6mm} p {4.3mm}|p {1.7mm} p {1.7mm} p {1.7mm} p {1.7mm} p {1.6mm} p {4.3mm}|}
		\toprule
		{Apps} & \multicolumn{6}{c|}{Line coverage} & \multicolumn{6}{c|}{Crashes} \\
		{} & {ST} & {EH} & {AP} & {TM} & {CB} & {CB$_{wd}$} & {ST} & {EH} & {AP} & {TM} & {CB} & {CB$_{wd}$} \\
		\midrule
			mileage & 38 & 23 & 58 & 40 & \cellcolor[HTML]{CDCAC7} 60 & 57 & 2 & 0 & \cellcolor[HTML]{CDCAC7} 15 & 9 & 4 & 4 \\
			bomber & 61 & 56 & 66 & \cellcolor[HTML]{CDCAC7} 97 & 88 & 87 & \cellcolor[HTML]{CDCAC7} 0 & \cellcolor[HTML]{CDCAC7} 0 & \cellcolor[HTML]{CDCAC7} 0 & \cellcolor[HTML]{CDCAC7} 0 & \cellcolor[HTML]{CDCAC7} 0 & \cellcolor[HTML]{CDCAC7} 0 \\
			mirrored & 31 & 16 & 38 & 46 & \cellcolor[HTML]{CDCAC7} 47 & \cellcolor[HTML]{CDCAC7} 47 & 0 & 0 & 0 & 1 & \cellcolor[HTML]{CDCAC7} 1 & \cellcolor[HTML]{CDCAC7} 1 \\
			batterydog & 59 & 5 & 72 & \cellcolor[HTML]{CDCAC7} 73 & 72 & 72 & 0 & 0 & 0 & \cellcolor[HTML]{CDCAC7} 1 & 0 & 0 \\
			triangle & 90 & \cellcolor[HTML]{CDCAC7} 91 & 90 & \cellcolor[HTML]{CDCAC7} 91 & \cellcolor[HTML]{CDCAC7} 91 & \cellcolor[HTML]{CDCAC7} 91 & 0 & 0 & 0 & 0 & \cellcolor[HTML]{CDCAC7} 1 & \cellcolor[HTML]{CDCAC7} 1 \\
			translate & 46 & 29 & 48 & 48 & \cellcolor[HTML]{CDCAC7} 49 & \cellcolor[HTML]{CDCAC7} 49 & \cellcolor[HTML]{CDCAC7} 1 & \cellcolor[HTML]{CDCAC7} 1 & \cellcolor[HTML]{CDCAC7} 1 & 0 & \cellcolor[HTML]{CDCAC7} 1 & \cellcolor[HTML]{CDCAC7} 1 \\
			anymemo & 26 & 18 & 50 & 42 & \cellcolor[HTML]{CDCAC7} 52 & 46 & 2 & 1 & 6 & 6 & \cellcolor[HTML]{CDCAC7} 7 & \cellcolor[HTML]{CDCAC7} 7 \\
			zooborns & 18 & 17 & 19 & 25 & \cellcolor[HTML]{CDCAC7} 26 & \cellcolor[HTML]{CDCAC7} 26 & \cellcolor[HTML]{CDCAC7} 3 & 0 & \cellcolor[HTML]{CDCAC7} 3 & \cellcolor[HTML]{CDCAC7} 3 & 1 & 1 \\
			qsettings & 40 & 23 & \cellcolor[HTML]{CDCAC7} 50 & 40 & 47 & 46 & 1 & 1 & 1 & 0 & \cellcolor[HTML]{CDCAC7} 1 & 0 \\
			wchart & 57 & 24 & 32 & 51 & \cellcolor[HTML]{CDCAC7} 85 & 83 & 2 & 1 & 0 & 0 & \cellcolor[HTML]{CDCAC7} 3 & \cellcolor[HTML]{CDCAC7} 3 \\
			addi & 17 & 16 & \cellcolor[HTML]{CDCAC7} 21 & 19 & 18 & 18 & 1 & 0 & \cellcolor[HTML]{CDCAC7} 8 & 1 & 3 & 3 \\
			LNM & 49 & 3 & 34 & 48 & \cellcolor[HTML]{CDCAC7} 50 & \cellcolor[HTML]{CDCAC7} 50 & 4 & 0 & 4 & \cellcolor[HTML]{CDCAC7} 7 & 2 & 1 \\
			gestures & 32 & 32 & 32 & 50 & \cellcolor[HTML]{CDCAC7} 78 & \cellcolor[HTML]{CDCAC7} 78 & \cellcolor[HTML]{CDCAC7} 0 & \cellcolor[HTML]{CDCAC7} 0 & \cellcolor[HTML]{CDCAC7} 0 & \cellcolor[HTML]{CDCAC7} 0 & \cellcolor[HTML]{CDCAC7} 0 & \cellcolor[HTML]{CDCAC7} 0 \\
			MNV & 35 & 13 & 64 & 42 & \cellcolor[HTML]{CDCAC7} 68 & \cellcolor[HTML]{CDCAC7} 68 & 2 & 1 & \cellcolor[HTML]{CDCAC7} 4 & \cellcolor[HTML]{CDCAC7} 4 & 1 & 1 \\
			wikipedia & 24 & 21 & 25 & \cellcolor[HTML]{CDCAC7} 31 & 19 & 19 & \cellcolor[HTML]{CDCAC7} 0 & \cellcolor[HTML]{CDCAC7} 0 & \cellcolor[HTML]{CDCAC7} 0 & \cellcolor[HTML]{CDCAC7} 0 & \cellcolor[HTML]{CDCAC7} 0 & \cellcolor[HTML]{CDCAC7} 0 \\
			dialer & 66 & 53 & 65 & 40 & \cellcolor[HTML]{CDCAC7} 73 & \cellcolor[HTML]{CDCAC7} 73 & 1 & 1 & 1 & \cellcolor[HTML]{CDCAC7} 3 & \cellcolor[HTML]{CDCAC7} 2 & \cellcolor[HTML]{CDCAC7} 2 \\
			photost & 24 & 9 & 26 & \cellcolor[HTML]{CDCAC7} 28 & 12 & 12 & 2 & 1 & 1 & \cellcolor[HTML]{CDCAC7} 3 & \cellcolor[HTML]{CDCAC7} 3 & \cellcolor[HTML]{CDCAC7} 3 \\
			battery & 92 & 55 & 55 & \cellcolor[HTML]{CDCAC7} 93 & 88 & 88 & 0 & 0 & 0 & \cellcolor[HTML]{CDCAC7} 3 & 0 & 0 \\
			aCal & 18 & 8 & 28 & \cellcolor[HTML]{CDCAC7} 29 & 22 & 19 & 3 & 0 & \cellcolor[HTML]{CDCAC7} 5 & 3 & 3 & 1 \\
			tomdroid & 55 & 24 & 57 & 53 & \cellcolor[HTML]{CDCAC7} 61 & 59 & 0 & 0 & \cellcolor[HTML]{CDCAC7} 4 & 0 & 2 & 2 \\
			RMP & 82 & 87 & 83 & 65 & \cellcolor[HTML]{CDCAC7} 92 & \cellcolor[HTML]{CDCAC7} 92 & 1 & 0 & 0 & 1 & \cellcolor[HTML]{CDCAC7} 2 & \cellcolor[HTML]{CDCAC7} 2 \\
			SpriteText & 62 & \cellcolor[HTML]{CDCAC7} 63 & 62 & \cellcolor[HTML]{CDCAC7} 63 & 61 & 59 & \cellcolor[HTML]{CDCAC7} 0 & \cellcolor[HTML]{CDCAC7} 0 & \cellcolor[HTML]{CDCAC7} 0 & \cellcolor[HTML]{CDCAC7} 0 & \cellcolor[HTML]{CDCAC7} 0 & \cellcolor[HTML]{CDCAC7} 0 \\
			LPG & 63 & 37 & \cellcolor[HTML]{CDCAC7} 89 & 82 & 0 & 0 & \cellcolor[HTML]{CDCAC7} 0 & \cellcolor[HTML]{CDCAC7} 0 & \cellcolor[HTML]{CDCAC7} 0 & \cellcolor[HTML]{CDCAC7} 0 & \cellcolor[HTML]{CDCAC7} 0 & \cellcolor[HTML]{CDCAC7} 0 \\
			ringdroid & 0 & 40 & 42 & 23 & \cellcolor[HTML]{CDCAC7} 47 & \cellcolor[HTML]{CDCAC7} 47 & 1 & 2 & \cellcolor[HTML]{CDCAC7} 4 & 2 & 2 & 2 \\
			sftp & 11 & 5 & 15 & 12 & \cellcolor[HTML]{CDCAC7} 18 & \cellcolor[HTML]{CDCAC7} 18 & 0 & 0 & 0 & 0 & \cellcolor[HTML]{CDCAC7} 3 & 1 \\
			PWMG & 3 & 6 & 7 & \cellcolor[HTML]{CDCAC7} 16 & 6 & 6 & 0 & 1 & 0 & 0 & \cellcolor[HTML]{CDCAC7} 2 & \cellcolor[HTML]{CDCAC7} 2 \\
			fbubble & 49 & 49 & 56 & \cellcolor[HTML]{CDCAC7} 82 & 74 & 72 & 0 & 0 & 0 & 0 & \cellcolor[HTML]{CDCAC7} 3 & \cellcolor[HTML]{CDCAC7} 3 \\
			myexp & 55 & 1 & 33 & 46 & \cellcolor[HTML]{CDCAC7} 65 & 63 & 0 & 0 & 0 & 1 & \cellcolor[HTML]{CDCAC7} 7 & \cellcolor[HTML]{CDCAC7} 7 \\
			sanity & 13 & 8 & 26 & 27 & \cellcolor[HTML]{CDCAC7} 36 & 35 & 1 & 0 & \cellcolor[HTML]{CDCAC7} 2 & 1 & \cellcolor[HTML]{CDCAC7} 2 & 1 \\
			SMT & \cellcolor[HTML]{CDCAC7} 87 & 2 & \cellcolor[HTML]{CDCAC7} 87 & 63 & \cellcolor[HTML]{CDCAC7} 87 & 85 & \cellcolor[HTML]{CDCAC7} 0 & \cellcolor[HTML]{CDCAC7} 0 & \cellcolor[HTML]{CDCAC7} 0 & \cellcolor[HTML]{CDCAC7} 0 & \cellcolor[HTML]{CDCAC7} 0 & \cellcolor[HTML]{CDCAC7} 0 \\
			alogcat & 65 & 33 & 73 & \cellcolor[HTML]{CDCAC7} 79 & 60 & 53 & 0 & 0 & 0 & 0 & \cellcolor[HTML]{CDCAC7} 2 & \cellcolor[HTML]{CDCAC7} 2 \\
			worldclock & 97 & 90 & \cellcolor[HTML]{CDCAC7} 98 & 94 & 95 & 95 & 1 & 1 & 0 & 1 & \cellcolor[HTML]{CDCAC7} 2 & \cellcolor[HTML]{CDCAC7} 2 \\
			mlife & 87 & 35 & 86 & 84 & \cellcolor[HTML]{CDCAC7} 92 & \cellcolor[HTML]{CDCAC7} 92 & 0 & 0 & 0 & 0 & \cellcolor[HTML]{CDCAC7} 2 & \cellcolor[HTML]{CDCAC7} 2 \\
			lbuilder & 22 & 28 & 28 & 26 & \cellcolor[HTML]{CDCAC7} 37 & 35 & 0 & 1 & 0 & 0 & \cellcolor[HTML]{CDCAC7} 4 & \cellcolor[HTML]{CDCAC7} 4 \\
			CDT & 63 & 31 & 65 & 85 & \cellcolor[HTML]{CDCAC7} 87 & \cellcolor[HTML]{CDCAC7} 87 & \cellcolor[HTML]{CDCAC7} 0 & \cellcolor[HTML]{CDCAC7} 0 & \cellcolor[HTML]{CDCAC7} 0 & \cellcolor[HTML]{CDCAC7} 0 & \cellcolor[HTML]{CDCAC7} 0 & \cellcolor[HTML]{CDCAC7} 0 \\
			bites & 26 & 15 & 42 & 36 & \cellcolor[HTML]{CDCAC7} 54 & \cellcolor[HTML]{CDCAC7} 54 & 2 & 0 & 5 & \cellcolor[HTML]{CDCAC7} 8 & 3 & 3 \\
			multisms & 40 & 26 & 74 & 57 & \cellcolor[HTML]{CDCAC7} 78 & \cellcolor[HTML]{CDCAC7} 78 & 0 & 1 & 0 & 1 & \cellcolor[HTML]{CDCAC7} 1 & \cellcolor[HTML]{CDCAC7} 1 \\
			yahtzee & \cellcolor[HTML]{CDCAC7} 69 & 3 & 46 & 6 & 51 & 46 & 1 & 0 & \cellcolor[HTML]{CDCAC7} 3 & 1 & \cellcolor[HTML]{CDCAC7} 3 & \cellcolor[HTML]{CDCAC7} 3 \\
			nectroid & 40 & 27 & 44 & 38 & \cellcolor[HTML]{CDCAC7} 46 & \cellcolor[HTML]{CDCAC7} 46 & 0 & 0 & 0 & \cellcolor[HTML]{CDCAC7} 2 & \cellcolor[HTML]{CDCAC7} 2 & \cellcolor[HTML]{CDCAC7} 2 \\
			anycut & 70 & 12 & \cellcolor[HTML]{CDCAC7} 71 & \cellcolor[HTML]{CDCAC7} 71 & 66 & 66 & 0 & 2 & 0 & 0 & \cellcolor[HTML]{CDCAC7} 3 & \cellcolor[HTML]{CDCAC7} 3 \\
			PMM & \cellcolor[HTML]{CDCAC7} 66 & 27 & 62 & 56 & 65 & 62 & 4 & 0 & \cellcolor[HTML]{CDCAC7} 11 & 3 & 4 & 4 \\
			manpages & 40 & 20 & 54 & 77 & \cellcolor[HTML]{CDCAC7} 78 & 74 & 0 & 0 & 0 & 1 & \cellcolor[HTML]{CDCAC7} 3 & \cellcolor[HTML]{CDCAC7} 3 \\
			zoffcc & 18 & 15 & 16 & \cellcolor[HTML]{CDCAC7} 20 & 16 & 16 & 3 & 0 & 4 & 1 & \cellcolor[HTML]{CDCAC7} 4 & \cellcolor[HTML]{CDCAC7} 4 \\
			amazed & 62 & 64 & 76 & 52 & \cellcolor[HTML]{CDCAC7} 84 & \cellcolor[HTML]{CDCAC7} 84 & 0 & 0 & \cellcolor[HTML]{CDCAC7} 1 & \cellcolor[HTML]{CDCAC7} 1 & \cellcolor[HTML]{CDCAC7} 1 & \cellcolor[HTML]{CDCAC7} 1 \\
			alarmclock & 72 & 15 & \cellcolor[HTML]{CDCAC7} 76 & 68 & 71 & 71 & \cellcolor[HTML]{CDCAC7} 6 & 0 & 4 & 4 & 5 & 5 \\
			hndroid & 13 & 5 & 11 & 8 & \cellcolor[HTML]{CDCAC7} 15 & \cellcolor[HTML]{CDCAC7} 15 & 0 & 1 & 0 & \cellcolor[HTML]{CDCAC7} 2 & \cellcolor[HTML]{CDCAC7} 2 & \cellcolor[HTML]{CDCAC7} 2 \\
			sboard & \cellcolor[HTML]{CDCAC7} 100 & 58 & \cellcolor[HTML]{CDCAC7} 100 & \cellcolor[HTML]{CDCAC7} 100 & \cellcolor[HTML]{CDCAC7} 100 & \cellcolor[HTML]{CDCAC7} 100 & \cellcolor[HTML]{CDCAC7} 0 & \cellcolor[HTML]{CDCAC7} 0 & \cellcolor[HTML]{CDCAC7} 0 & \cellcolor[HTML]{CDCAC7} 0 & \cellcolor[HTML]{CDCAC7} 0 & \cellcolor[HTML]{CDCAC7} 0 \\
			hotdeath & 16 & 63 & 73 & 75 & \cellcolor[HTML]{CDCAC7} 80 & 76 & 1 & 3 & 2 & 0 & \cellcolor[HTML]{CDCAC7} 5 & \cellcolor[HTML]{CDCAC7} 5 \\
			dalvik-exp & 23 & 6 & \cellcolor[HTML]{CDCAC7} 72 & 70 & 64 & 64 & 1 & 0 & \cellcolor[HTML]{CDCAC7} 5 & 3 & 4 & 4 \\
			jamendo & 10 & 13 & 28 & 9 & \cellcolor[HTML]{CDCAC7} 30 & \cellcolor[HTML]{CDCAC7} 30 & \cellcolor[HTML]{CDCAC7} 5 & 3 & 0 & 0 & \cellcolor[HTML]{CDCAC7} 5 & \cellcolor[HTML]{CDCAC7} 5 \\
			importcont & 57 & 2 & 53 & 42 & \cellcolor[HTML]{CDCAC7} 78 & 74 & 0 & 0 & 0 & 0 & \cellcolor[HTML]{CDCAC7} 1 & \cellcolor[HTML]{CDCAC7} 1 \\
			blokish & 36 & 35 & 49 & \cellcolor[HTML]{CDCAC7} 52 & 45 & 45 & 0 & 0 & \cellcolor[HTML]{CDCAC7} 2 & 0 & \cellcolor[HTML]{CDCAC7} 2 & \cellcolor[HTML]{CDCAC7} 2 \\
			Book-cat & 4 & 4 & 33 & 35 & \cellcolor[HTML]{CDCAC7} 38 & \cellcolor[HTML]{CDCAC7} 38 & 0 & 1 & 2 & 4 & \cellcolor[HTML]{CDCAC7} 4 & 0 \\
			Templaro & 55 & 76 & \cellcolor[HTML]{CDCAC7} 87 & 60 & 86 & 83 & 0 & 1 & 0 & 2 & \cellcolor[HTML]{CDCAC7} 3 & \cellcolor[HTML]{CDCAC7} 3 \\
			DAC & 53 & 48 & 76 & 88 & \cellcolor[HTML]{CDCAC7} 94 & 91 & \cellcolor[HTML]{CDCAC7} 0 & \cellcolor[HTML]{CDCAC7} 0 & \cellcolor[HTML]{CDCAC7} 0 & \cellcolor[HTML]{CDCAC7} 0 & \cellcolor[HTML]{CDCAC7} 0 & \cellcolor[HTML]{CDCAC7} 0 \\
			Agrep & 37 & 8 & 58 & \cellcolor[HTML]{CDCAC7} 63 & 61 & 58 & 0 & 0 & \cellcolor[HTML]{CDCAC7} 7 & 2 & \cellcolor[HTML]{CDCAC7} 7 & \cellcolor[HTML]{CDCAC7} 7 \\
			Syncmypix & 15 & 18 & 21 & 25 & \cellcolor[HTML]{CDCAC7} 26 & \cellcolor[HTML]{CDCAC7} 26 & 1 & 1 & 0 & 1 & \cellcolor[HTML]{CDCAC7} 3 & \cellcolor[HTML]{CDCAC7} 3 \\
			tippytipper & 72 & 9 & 86 & 84 & \cellcolor[HTML]{CDCAC7} 89 & \cellcolor[HTML]{CDCAC7} 89 & 0 & 0 & 0 & 0 & \cellcolor[HTML]{CDCAC7} 2 & \cellcolor[HTML]{CDCAC7} 2 \\
			WHAMS & \cellcolor[HTML]{CDCAC7} 80 & 0 & 77 & 69 & 79 & 79 & 0 & 0 & 0 & \cellcolor[HTML]{CDCAC7} 1 & \cellcolor[HTML]{CDCAC7} 1 & \cellcolor[HTML]{CDCAC7} 1 \\
			A2dp & 29 & 14 & 40 & 45 & \cellcolor[HTML]{CDCAC7} 47 & 42 & \cellcolor[HTML]{CDCAC7} 6 & 0 & \cellcolor[HTML]{CDCAC7} 6 & 0 & 3 & 3 \\
			\midrule
			Avg/Sum & 46 & 27 & 53 & 52 & \cellcolor[HTML]{CDCAC7} 58 & 57 & 58 & 25 & 111 & 87 & \cellcolor[HTML]{CDCAC7} 137 & 126 \\
		\bottomrule
	\end{tabular}
	\vspace{-1mm}
	\caption{Coverage and the number of crashes reported by all the tools in the \androtest dataset.
	ST: \stoat, EH: \ehbdroid, AP: \ape, TM: \timemachine, CB: \tool, CB\textsubscript{wd}: \tool without dependency feedback}
	\label{tbl:crash_and_coverage_in_androtest}
	\vspace{1mm}
\end{table}

\subsubsection{Performance on benchmark apps}
To investigate how our technique performs with respect to prior work, we use the \androtest benchmark apps. Specifically, we compared the achieved code coverage and the number of crashes found by \tool with the state-of-the-art model-based techniques \stoat~\cite{stoat} and \ape~\cite{ape}, checkpoint-based technique \timemachine~\cite{timemachine}, and callback-driven technique \ehbdroid~\cite{ehbdroid}.
Unfortunately, we could not make the publicly available version of \ehbdroid work on our test apps due to the incompatibility of their instrumentation module with our test subjects.
Instead, we implemented their testing strategies by modifying \tool in three ways: \textbf{(i)} we consider only those $58$ callbacks supported by \ehbdroid, 
\textbf{(ii)} we disabled dependency and crash guidance, and
\textbf{(iii)} we restricted primitive argument values to those used by \ehbdroid instead of the values computed by our argument generation module.

In \Tbl{tbl:crash_and_coverage_in_androtest}, we present the statement coverage achieved as well as the crashes triggered by all tools on the benchmark apps.
\looseness=-1

\begin{figure}[t]
	\includegraphics[width=0.9\columnwidth]{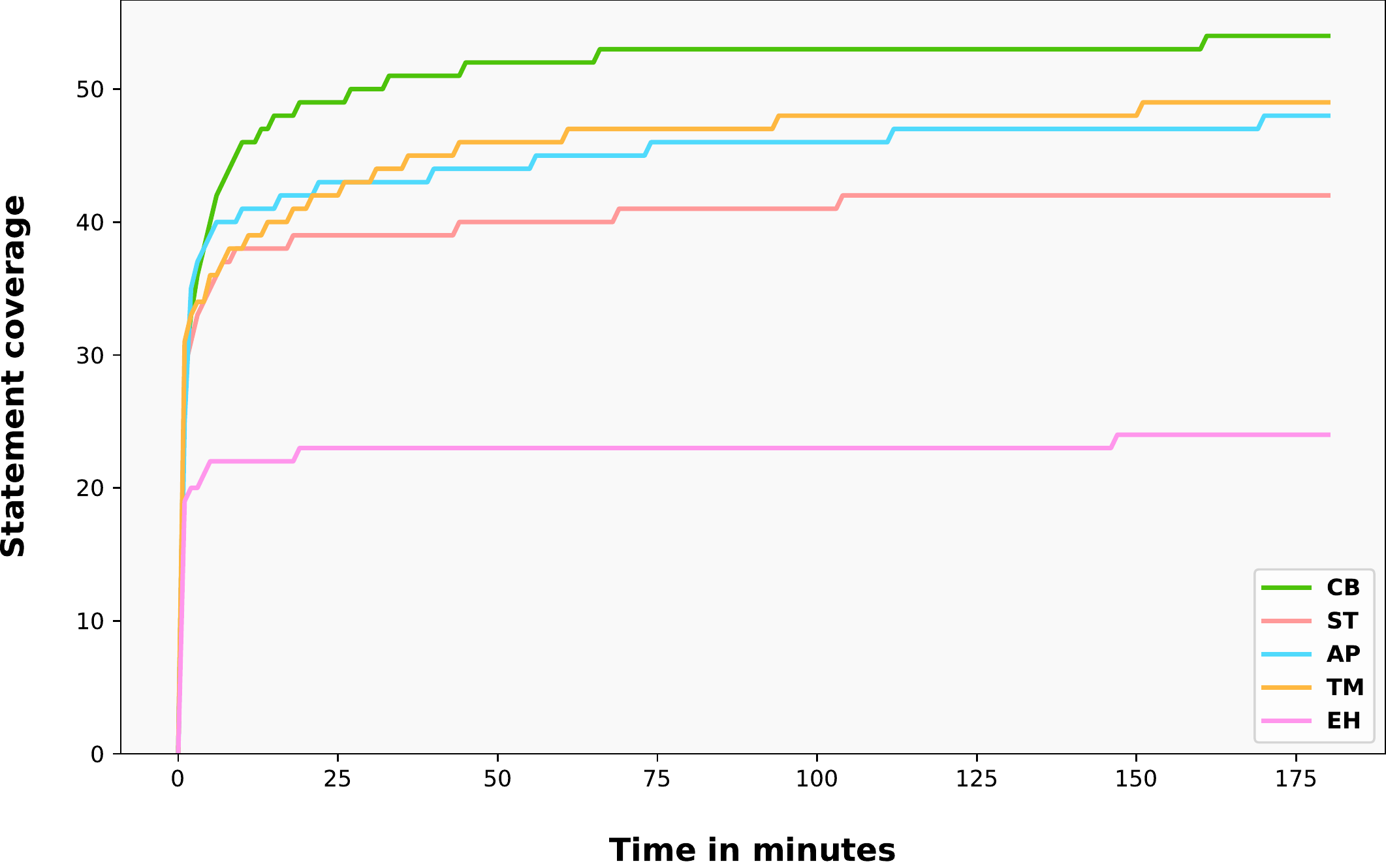}
	\centering
	\caption{Progression of coverage over time by all the tools on the \androtest dataset.
		Tool codes are similar to \Tbl{tbl:crash_and_coverage_in_androtest}}
	\label{fig:coverage_over_time}
	\vspace{1.5mm}
\end{figure}

\mypar{Coverage.}
\begin{figure}[t]
	\captionsetup[subfigure]{font=small}
	\centering
	\begin{subfigure}[b]{0.22\columnwidth}
		\centering
		\includegraphics[width=\textwidth]{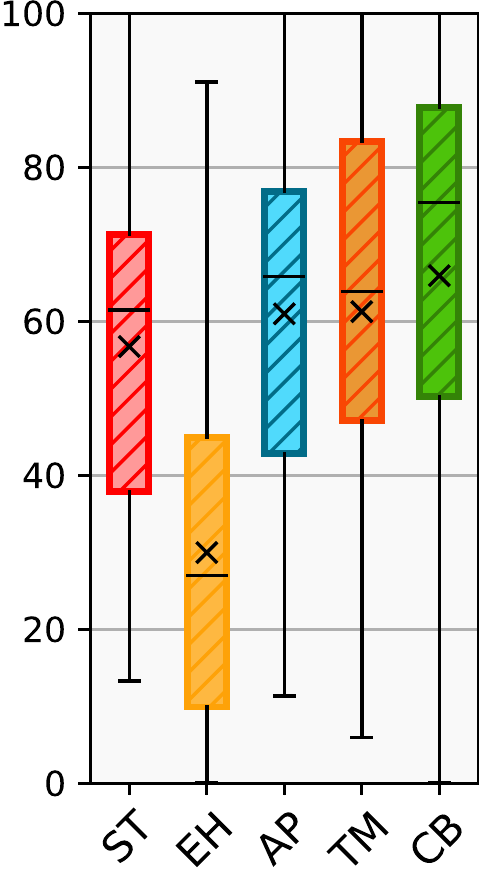}
		\caption{$<$1K ($\numappslessthanonek$)}
		\label{fig:coverage_boxplot_lt_1k}
	\end{subfigure}
	\hfill
	\begin{subfigure}[b]{0.22\columnwidth}
		\centering
		\includegraphics[width=\textwidth]{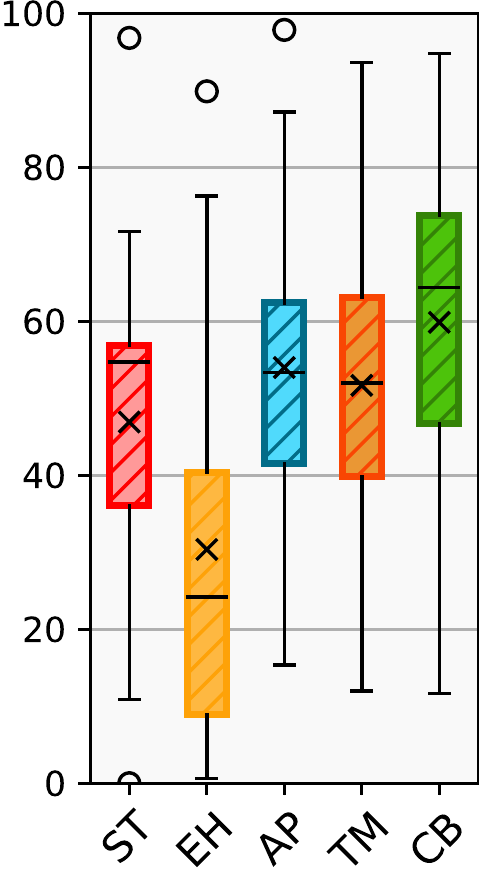}
		\caption{[1K,3K) ($\numappsgreaterthanoneklessthanthreek$)}
		\label{fig:coverage_boxplot_gt_1k_lt_3k}
	\end{subfigure}
	\hfill
	\begin{subfigure}[b]{0.22\columnwidth}
		\centering
		\includegraphics[width=\textwidth]{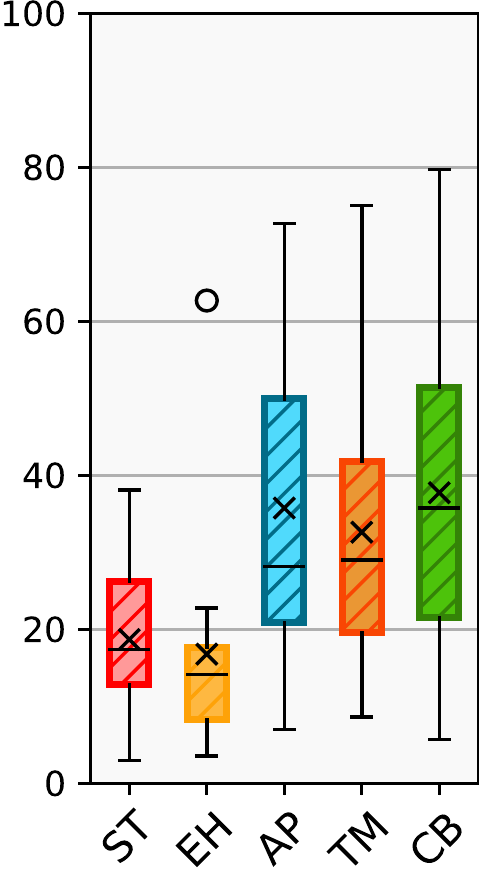}
		\caption{$\geq$3K ($\numappsgreaterthanthreek$)}
		\label{fig:coverage_boxplot_gt_3k}
	\end{subfigure}
	\hfill
	\begin{subfigure}[b]{0.22\columnwidth}
		\centering
		\includegraphics[width=\textwidth]{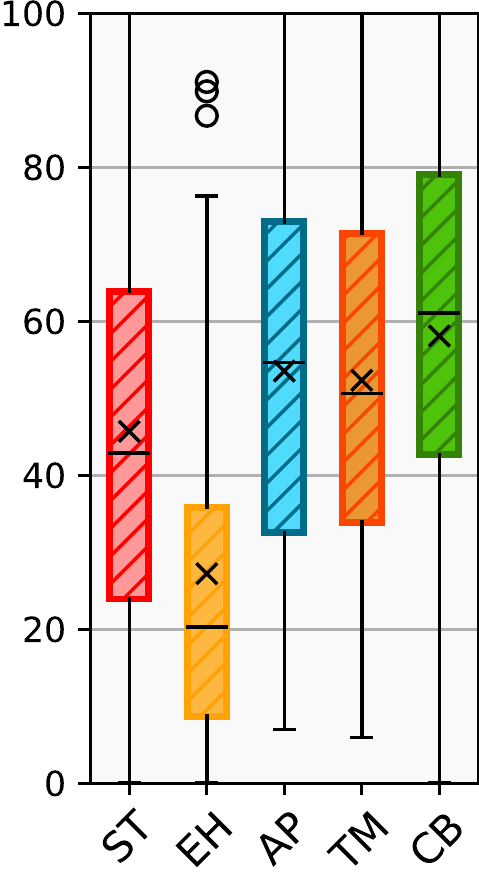}
		\caption{all ($\numandrotestappsworked$)}
		\label{fig:coverage_boxplot_all}
	\end{subfigure}
	\caption{Coverage (Y-axis) achieved on \androtest, grouped by app size (Lines of Code).
		Number of apps in a size group is indicated in parentheses.
		`x' denotes the mean of a boxplot}
	\label{fig:coverage_boxplot}
	\vspace{2mm}
\end{figure}
We find that \tool achieves higher code coverage than \stoat, \ehbdroid, \ape and \timemachine for $\numappswithcoveragemorethanstoat$, $\numappswithcoveragemorethanehbdroid$, $\numappswithcoveragemorethanape$, and $\numappswithcoveragemorethantimemachine$ apps, respectively.
Moreover, \tool achieves the best coverage in $\columbuscoveragebestinapps$ apps, followed by \timemachine ($\timemachinecoveragebestinapps$ apps), \ape ($\apecoveragebestinapps$ apps), \stoat ($\stoatcoveragebestinapps$ apps), and \ehbdroid ($\ehbdroidcoveragebestinapps$ apps).
To gain an overall view of the tools' performances, we report the average code coverage, achieved by each tool across all apps, in the last row of \Tbl{tbl:crash_and_coverage_in_androtest}.
As can be seen, \tool attains the highest ($\columbusaveragecoverage\%$) coverage on average, followed by \ape ($\apeaveragecoverage\%$), \timemachine ($\timemachineaveragecoverage\%$), \stoat ($\stoataveragecoverage\%$), and \ehbdroid ($\ehbdroidaveragecoverage\%$).
\Fig{fig:coverage_over_time} shows the progression of coverage over time for all the tools averaged across all the benchmark apps.
Starting from the $\timeaftercolumbuscoverageexceedsothers$th minute, the coverage achieved by \tool exceeds other tools.
Until approximately the $\timeuntilcoverageincreasesfast$th minute, the coverage increases at a fairly fast rate, after that, it starts to slow down.
Further, the boxplot in \Fig{fig:coverage_boxplot} shows the \textit{spread} of the coverage achieved by all the tools grouped by the size of the apps.
We use group sizes identical to the ones used in previous work~\cite{timemachine}.
As the figure shows, \tool exhibits significant improvement over other tools in terms of coverage for all size groups.

The improvement in coverage for \tool can be attributed to its systematic exploration of the callbacks.
While UI-based techniques struggle to generate complex events and appropriate user input, \tool sidesteps this problem by directly calling the callbacks and supplying argument values (computed by the argument generation module) that are likely to explore additional code paths.
In addition, the crash-guidance feedback helps \tool to make the best use of the time-budget by preventing the exploration from getting stuck at individual crashes for a long time.

\Fig{fig:randommusicplayer} shows a code snippet from the \code{RandomMusicPlayer} app from \androtest.
This example shows an interesting case where \tool naturally enjoys clear benefits over previous, more  ``heavyweight'' techniques that use symbolic execution~\cite{Anand:2012}, and other UI-testing tools.
To explore all the branches (\code{if} conditions), a UI-based tool would need to \code{click} on all corresponding buttons, which is challenging.
\acteve~\cite{Anand:2012} solves this problem by concolically executing the app together with an instrumented version of the \android framework.
Since, in our case, \tool introspects the app heap to retrieve live objects, we observed the coverage of this app quickly going up, because \tool invokes the \code{onClick} callback with all the button \code{Views} already present in the heap.
\looseness=-1

\begin{figure}[t]
	\vspace{-2mm}
\begin{lstlisting}[
	language=java,
	xleftmargin=10pt,
	xrightmargin=10pt,
	frame=single,
	caption=,
	basicstyle=\tiny,
	stepnumber=1,
	firstnumber=1]
public void onClick(View target) {
  // Send intent according to the button clicked
  if (target == mPlayButton) {
     startService(new Intent(MusicService.ACTION_PLAY));
  } else if (target == mPauseButton) {
     startService(new Intent(MusicService.ACTION_PAUSE));
  } else if (target == mSkipButton) {
     startService(new Intent(MusicService.ACTION_SKIP));
  } else if (target == mRewindButton) {
    startService(new Intent(MusicService.ACTION_REWIND));
  } else if (target == mStopButton) {
     startService(new Intent(MusicService.ACTION_STOP));
  } else if (target == mEjectButton) {
     showUrlDialog();
  }
}
\end{lstlisting}
	\vspace{-2mm}
	\caption{Code snippet (redacted) from RandomMusicPlayer}
	\label{fig:randommusicplayer}
	\vspace{2mm}
\end{figure}

To better understand the challenges \tool faces during exploration, we manually examined $\numappsanalyzedwherecolumbusisworse$ of those apps where \tool did not achieve the best coverage.
We summarize our findings next: \textbf{(i)} For callbacks where the symbolic execution timed out, the \moduleargumentgeneration module could not return any useful value.
As a result, \tool fell back to its default strategy of trying out random argument values, which negatively affected the coverage.
\textbf{(ii)} There exist callbacks that are \textit{stateful}.
That is, the application logic is conditioned on \code{class} variables.
Note that \tool is not state-aware, therefore this challenge is orthogonal to what \tool aims to solve.
\textbf{(iii)} For unconstrained callback arguments, we use random values from a predefined list, which might be ineffective.
For instance, the \code{yahtzee} app lists the game moves in a drop-down list.
A move can be chosen by the \code{arg2} argument (unconstrained) of the \code{onItemSelected(\_, \_, arg2, \_)} callback, which then looks up the appropriate UI object using that argument.
Many such values of \code{arg2} that we supply could be invalid, while UI-based techniques can ``blindly'' \code{click} on the list item without being aware of the valid values of that argument.

\mypar{Crashes.}
\tool found a total of $\totalandrotestcrashwithFP$ crashes. After excluding the potential false positives, the total number of crashes become $\totalandrotestcrash$ (\Tbl{tbl:crash_and_coverage_in_androtest}). As presented in \Tbl{tbl:crash_found_by_us}, \tool found crashes of $\typesofandrotestcrash$ different types 
in $\numandrotestappswithcrash$ out of $\numandrotestappsworked$ apps in the \androtest dataset.
Compared to \stoat, \ehbdroid, \ape, and \timemachine, \tool discovered $\FPeval{\v}{round(\numcrashescolumbus/\numcrashesstoat, 2)}\v$, $\FPeval{\v}{round(\numcrashescolumbus/\numcrashesehbdroid, 2)}\v$, $\FPeval{\v}{round(\numcrashescolumbus/\numcrashesape, 2)}\v$, and $\FPeval{\v}{round(\numcrashescolumbus/\numcrashestimemachine, 2)}\v$ times more crashes, respectively.
To acquire a better understanding of how the tools perform on individual apps, we calculated the number of apps for which each tool discovers the most number of crashes.
While \stoat, \ehbdroid, \ape, and \timemachine finds the most crashes in  $\stoatcrashesbestinapps$, $\ehbdroidcrashesbestinapps$, $\apecrashesbestinapps$ , and $\timemachinecrashesbestinapps$ apps, respectively, \tool performs the best for the highest ($\columbuscrashesbestinapps$) number of apps.
\looseness=-1

\mypar{False positive analysis.}
Our strategy of invoking callbacks directly, sometimes with artificially-prepared arguments, can potentially lead to false positives (FP), \ie, generate spurious crashes that cannot be triggered when the app is normally exercised from the UI.
Since \stoat, \ape, and \timemachine are UI-driven testing tools, they always generate legitimate crashes.
For \tool, we identify two potential reasons for FPs and quantify their prevalence.

\textbf{(i)} \textbf{Disabled UI elements.}
Since \tool does not access the UI state of the app, it may (incorrectly) invoke a callback $cb_d$ associated with a widget $W$, which is disabled at the time of invocation.
If such a callback $cb_d$ exists in an app, then there exists another callback $cb_e$ that calls \code{W.setEnabled()} to enable the widget.
We found that only $\numappcallbackswithsetenabled$ ($cb_e$) out of $\num{\numappcallbacks}$ callbacks in our benchmark apps contain such calls.
Now, \code{setEnabled} calls from inside the lifecycle callbacks are not problematic.
Because, the latter is called by the \android framework, which enables the respective UI elements as part of the initialization of the app.
Among those $\numappcallbackswithsetenabled$, only $\numappcallbackswithsetenablednonlifecycle$ callbacks are non-lifecycle ones, which is negligible with respect to the total number of callbacks.

\textbf{(ii)} \textbf{Uninitialized nested object argument.}
If a callback expects an object argument of \code{class A} that we do not find in the heap, we create an instance $a$ by invoking the class constructor $C$.
However, instances created in this way may be partially uninitialized.
Suppose, $A$ contains a field $A.b$ of \code{class B}, which $C$ leaves uninitialized.
If the callback attempts to access $A.b$, then it will result in a \npe.
This is a spurious crash, because when the app is exercised from the UI, the framework would invoke the callback with a correctly constructed object.
In case of the benchmark apps, we needed to create object arguments for only $\numappcallbacksinvokedwithreflection$ ($\FPeval{\v}{round(\numappcallbacksinvokedwithreflection/\numappcallbacks*100, 2)}\v\%$) out of $\num{\numappcallbacks}$ callbacks.
Unfortunately, there is no straightforward way to estimate further how many of these callbacks require nested object arguments.
Even then, since we already invoke object creation for a reasonably small number of callbacks, that makes the probability of such FPs minimal.

To investigate into our potential sources of FPs, we first collected all $\numcolumbusonlycrash$ crashes that are found only by \tool, but not by any of those tools.
Then, we manually verified those reports to determine potential FPs.
We call a report \textit{legitimate}, if we can reproduce a crash with the same stack trace by exercising the app from the UI.
To do that, we collected a sequence of callback invoked immediately before the crash from our tool's output log, and also reviewed the relevant part of the source code to seek further guidance.
If we failed to reproduce the crash within a reasonable number of tries, we flagged the report as FP.
Note that, this estimate is conservative and best-effort, because it includes true crash reports that we could not reproduce because of \android apps' inherent statefulness.
At the end, we failed to reproduce $\numcolumbusonlycrashnotreproduced$ crashes out of total $\totalandrotestcrashwithFP$ crashes, which, even in the worst case, translates to a mere $\FPeval{\v}{round(\numcolumbusonlycrashnotreproduced/\numcrashescolumbuswithFP*100, 2)}\v\%$ FP rate.
We argue that this amount of FPs is acceptable in practice, given the benefits (extra crashes, coverage) that our approach brings.
\looseness=-1

\vspace{1mm}
\begin{mdframed}[style=graybox]
	\textbf{RQ1}: Compared to the state-of-the-art tools, \tool attains the highest coverage on average ($\columbusaveragecoverage\%$), and discovers the most number of crashes ($\numcrashescolumbus$) on the \androtest dataset. 
\end{mdframed}

\subsubsection{Performance on real-world apps}
\begin{table}[t]
	\tiny
	\begin{minipage}[t]{.30\columnwidth}
		
	\begin{tabular}{lr}
		\toprule
		Category & Count \\
		\midrule
		Education &  27 \\
		Games & 26 \\
		Personalization &  18 \\
		Tools &  17 \\
		Multimedia &  11 \\
		Photography &   4 \\
		Lifestyle &   7 \\
		Health \& Fitness &   4 \\
		Food \& Drink &   4 \\
		Entertainment &   6 \\
		Travel \& Local &   6 \\
		Business &   2 \\
		Productivity &   4 \\
		Others &   4 \\
		\midrule
		Total & $\numrealworldapps$ \\
		\bottomrule
	\end{tabular}
	\vspace{1mm}
	\captionsetup{justification=raggedright}
	\caption{Real-world app categories}
	\label{tbl:app_types_real_world}

	\end{minipage}
	\hfill
	\begin{minipage}[t]{.65\columnwidth}
			\begin{tabular}{rlrr}
	\toprule
	{ID} & {Exception type} & {A} & R\\
	\midrule
	 $1$ & NullPointerException & $52$ & $22$ \\
     $2$ & IllegalStateException & $16$ & $26$\\
	 $3$ & ArrayIndexOutOfBoundsException & $7$ & $4$\\
	 $4$ & IndexOutOfBoundsException & $10$ & $2$\\
	 $6$ & CursorIndexOutOfBoundsException & $10$ & - \\
	 $7$ & UnsatisfiedLinkError & $6$ & - \\
	 $8$ & RuntimeException & $1$ & $2$\\
	 $9$ & IllegalArgumentException & $15$ & $4$ \\
	 $10$& ClassCastException & $1$ & $2$\\
	 $12$ & StaleDataException & $3$ & - \\
	 $13$ & ActivityNotFoundException & $8$ & $6$ \\
	 $14$ & SQLiteDoneException & $1$ & - \\
	 $15$ & NumberFormatException & $1$ & - \\
	 $16$ & App Exceptions & $6$ & $2$ \\
	 \midrule
	 & Total & $\totalandrotestcrash$ & $\totalrealworldcrash$ \\
	 \bottomrule
	\end{tabular}
	\vspace{1mm}
	\caption{Crashes found by \tool.
	A: \androtest, R: Real-world dataset}
	\label{tbl:crash_found_by_us}

	\end{minipage}
	\vspace{2mm}
\end{table}
To understand the practicality of our approach, we tested \tool on the real-world dataset.
In line with the previous approaches~\cite{sapienz,stoat,timemachine}, we only considered the number of crashes discovered by our tool for this evaluation.
\looseness=-1

\mypar{Crashes.}
As shown in \Tbl{tbl:crash_found_by_us}, we discovered a total of $\num{\totalrealworldcrash}$ crashes of $\typesofrealworldcrash$ different types 
in $\numrealworldappswithcrash$ out of $\numrealworldapps$ apps, where \ise ($\FPeval{\v}{round(\illegalstateexceptionsinrealworldapps/\totalrealworldcrash*100, 2)}\v\%$) and \npe ($\FPeval{\v}{round(\nullpointerexceptionsinrealworldapps/\totalrealworldcrash*100, 2)}\v\%$) are the most prevalent ones.

\vspace{1mm}
\begin{mdframed}[style=graybox]
	\textbf{RQ2}: \tool is able to find $\num{\totalrealworldcrash}$ crashes in $\numrealworldappswithcrash$ out of $\numrealworldapps$ real-world Play Store apps, belonging to $\numrealworldappcategories$ categories.
\end{mdframed}

\subsubsection{Effectiveness of dependency feedback}
To show the effectiveness of the dependency feedback, we performed an ablation study by comparing \tool with \tool\textsubscript{wd}, a modified version of our tool 
that runs without the dependency feedback.
\Tbl{tbl:crash_and_coverage_in_androtest} presents the results of this experiment on the \androtest dataset.

While the coverage attained by both \tool and \tool~\textsubscript{wd} are comparable, the latter finds $\FPeval{\v}{round(\totalandrotestcrash-\numcrashescolumbusdependency, 0)}\v$ fewer crashes than the former in $\numappsaffectedbydependencyfeedback$ apps.
By manually inspecting those apps---\code{Book-cat}, \code{qsettings}, \code{sanity}, \code{sftp}, and \code{aCal}, we can confirm that the additional crashes are correlated with the number of dependency relations discovered.
In other words, due to higher than average ($\FPeval{\v}{round(\totaldependencyrelations/\numandrotestappsworked, 0)}\v$ dependencies/app) number of dependencies being present in those apps, the dependency feedback could indeed help \tool in triggering more crashes.
In addition, \tool achieved better coverage than any other tool for the first four apps.

\vspace{1mm}
\begin{mdframed}[style=graybox]
	\textbf{RQ3}: The dependency feedback used by \tool is useful for triggering crashes in apps, particularly for those ones with large amount of inter-callback dependencies.
\end{mdframed}

\section{Limitations}
\label{sec:limitation}

\mypar{Inferring correct value of the object fields.}
Currently, our \moduleargumentgeneration module can only infer the correct values of the primitive arguments.
However, it can be extended to support object arguments as well.
Consider the callback: \code{onKeyDown (int keyCode, KeyEvent event)}, which gets called when a key down event occurs.
Now, \code{event.getUnicodeChar()} API returns the Unicode character $c$ generated by that key event.
If a callback has paths conditioned on $c$, we can infer its correct values by symbolizing the return value of the API.
The inferred values can be used during testing to either dynamically set the correct value of the appropriate field of the \code{event} argument, or `hook' the \code{getUnicodeChar()} API to alter its return value---exercising more paths in effect.

\mypar{Creating values for login.}
There are \android apps which requires a userid and password to login first before one can explore its functionality.
\tool in its current shape can not detect such a login prompt, and enter the username and password automatically to explore such an app.
However, this is a limitation that we share with the existing state-of-the art tools, and an interesting direction for future work.

\section{Related Work}

\mypar{Random.}
Random testing based techniques such as \monkey~\cite{monkey} delivers random events.
\dynodroid~\cite{dynodroid}, in addition, considers system-level events, and monitors which events have registered listeners in the app to prioritize certain events depending on the context.
\puma~\cite{hao2014puma} presents an automation framework that has support for custom dynamic exploration strategies.
However, random testing strategies, though popular, often get stuck in a ``local optima,'' making no further progress.

\mypar{Model-based.}
Model-based testing approaches guide the exploration of the app by deriving a model of the app's UI.
Though some techniques require this model to be provided manually~\cite{White:2000, Yuan:2010, Merwe:2012}, 
others reconstruct the UI model using dynamic app exploration~\cite{Li:2017droidbot, Su:2022, stoat, Borges:2018, Amalfitano:2015, Memon:2003}.
Other techniques also perform model abstraction via identifying the structural similarities between different layouts~\cite{triggerscope}, model refinement by merging several UI interaction~\cite{ape}, and state recovery using snapshotting~\cite{timemachine}.
Model based testing techniques oftentimes suffer from state explosion if there are too many states in the app.
Therefore, they need to strike a balance between model completeness and scalability.

\mypar{Symbolic execution-based.}
Anand \etal~\cite{Anand:2012} concolically executes both the \android framework and the entire app, which is precise, but not scalable.
In contrast, \tool does symbolic execution only within a callback to strike a balance between precision, and scalability.
Another approach~\cite{Jensen:2013} starts the symbolic exploration in reversed order from the target blocks, and obtains the sequences of events to reach these targets.
Additionally, several other techniques were introduced for the symbolic execution of the apps that include libraries as well~\cite{Mirzaei:2012, Gao:2018}. %

\mypar{Hybrid.}
Similar to \tool, several approaches also employ hybrid techniques, \ie, combination of static and dynamic strategies, for app exploration.
In particular, \cite{Azim:2013, Lai:2019, Yang:2013, Yan:2020, Guo:2020} reconstruct the app model statically, followed by dynamic exploration.
Other techniques use static analysis to discover dependencies between different application components, and use it during the dynamic exploration~\cite{Azim:2013, Wang:2020combodroid, Jensen:2013, Sadeghi:2017, Guo:2020, Arlt:2012}.
Another guided exploration technique CAR~\cite{Wong22} uses a static constraint analysis to keep the symbolic execution scalable and obviate the need for whole program symbolic execution.
In contrast, \tool aims to maximize coverage similar to other app testing tools limiting the scope of the symbolic execution only within the callback and sets up the environment in an under-constrained manner.
Moreover, during the dynamic exploration, \tool uses a type-guided object matching to supply an existing, well-formed object to the callback.
Whereas, CAR resorts to a refinement-based construction of heap objects, guided by a crash-oracle.
A crash resulting from a malformed object acts as a ‘hint’ to fix the shape of the object. 
\ehbdroid~\cite{ehbdroid} instruments the app statically to include callback invocations within the app code in order to invoke them directly. 
However, their technique is not generic, and suffers from limitations as discussed before.

\section{Conclusion}
\label{sec:conclusion}
This paper proposed \tool, a callback-driven \android app testing technique that improves over the state-of-the-art in three aspects:
\textbf{(i)} systematically identifying the callbacks present in an app,
\textbf{(ii)} inferring coverage maximizing primitive arguments, while generating object arguments in an \android API-agnostic manner,
and \textbf{(iii)} providing \feedbackdatadependency and \feedbackcrashguidance as `feedback' to increase the probability of triggering uninitialized data related crashes, and preventing the tool from rediscovering same bugs, respectively.
In our evaluation, \tool outperformed state-of-the-art model-driven, checkpoint-based, and callback-driven testing tools both in terms of crashes and coverage.

\section{Acknowledgments}
\label{sec:acknowledgement}
We want to thank our anonymous reviewers for their valuable comments and feedback to improve our paper.
This research is supported in part by DARPA under the agreement number N66001-22-2-4037, by the NSF under award $2107101$, Google ASPIRE Award, and by the Dutch Ministry of Economic Affairs and Climate Policy (EZK) through the AVR project.
The U.S. Government is authorized to reproduce, and distribute reprints for Governmental purposes notwithstanding any copyright notation thereon.
The views and conclusions contained herein are those of the authors, and should not be
interpreted as necessarily representing the official policies or endorsements, either expressed or implied, of DARPA or the U.S. Government.

\bibliographystyle{plain}
\bibliography{bibs/biblio}
\end{document}